\documentclass[11pt,a4paper]{article}
\usepackage{jheppub}

\usepackage{epsfig,multicol}
\usepackage{amsfonts}
\usepackage{multirow}
\usepackage{mathrsfs}
\usepackage{graphicx}
\usepackage{amsmath}
\usepackage{amssymb}
\usepackage{bm}
\usepackage{bbm}
\usepackage{color}
\usepackage{array}
\usepackage{diagbox}

\def\pslash{p\!\!\!\slash}
\def\pbarslash{\bar{p}\!\!\!\slash}

\def\Pslash{P\!\!\!\slash}

\def\OMIT#1{}

\newcommand{\nn}{\nonumber}

\newcommand{\beq}{\begin{equation}}
\newcommand{\eeq}{\end{equation}}
\newcommand{\bqa}{\begin{eqnarray}}
\newcommand{\eqa}{\end{eqnarray}}

\newcommand\fverb{\setbox\fverbbox=\hbox\bgroup\verb}
\newcommand\fverbdo{\egroup\medskip\noindent%
			\fbox{\unhbox\fverbbox}\ }
\newcommand\fverbit{\egroup\item[\fbox{\unhbox\fverbbox}]}
\newbox\fverbbox

\makeatletter

\newcommand{\Rmnum}[1]{\expandafter\@slowromancap\romannumeral #1@}
\makeatother

\title{\mbox{}\\[10pt]
Next-to-leading-order QCD corrections to a vector bottomonium radiative decay into a charmonium
}
\author[a]{Yu-Dong Zhang,}
\author[b,c]{Feng Feng,}
\author[a]{Wen-Long Sang,}
\author[d]{Hong-Fei Zhang}

\affiliation[a]{School of Physical Science and Technology, Southwest University, Chongqing 400700, China}
\affiliation[b]{China University of Mining and Technology, Beijing 100083, China}
\affiliation[c]{Institute of High Energy Physics and Theoretical Physics Center for Science Facilities, Chinese Academy of Sciences, Beijing 100049, China}
\affiliation[d]{College of Big Data Statistics, Guizhou University of Finance and Economics, Guiyang, 550025, China}

\emailAdd{zhangyudong@email.swu.edu.cn}
\emailAdd{F.Feng@outlook.com}
\emailAdd{wlsang@swu.edu.cn:corresponding author}
\emailAdd{shckm2686@163.com}

\date{\today}
\abstract{
Within the  framework of nonrelativistic QCD (NRQCD) factorization, we calculate the next-to-leading-order (NLO)
perturbative corrections to the radiative decay $\Upsilon\to \eta_c(\chi_{cJ})+\gamma$.
Both the helicity amplitudes and the helicity decay widths are obtained.
It is the first computation for the processes involving both bottomonium and charmonium  at two-loop accuracy.
By employing the Cheng-Wu theorem, we are able to convert most of complex-valued master integrals (MIs) into real-valued MIs, which makes the numerical integration much efficient.
Our results indicate the $\mathcal{O}(\alpha_s)$ corrections are moderate for $\eta_c$ and $\chi_{c2}$ production,
and are quite marginal for $\chi_{c0}$ and $\chi_{c1}$ production.
It is impressive to note the NLO corrections considerably reduce the renormalization scale dependence in both the decay widths and the
branching fractions for $\chi_{cJ}$, and slightly improve that for $\eta_c$.
With the NRQCD matrix elements evaluated via the Buchm\"uller-Tye potential model, we find the decay width for $\eta_c$ production is one-order-of-magnitude larger than $\chi_{cJ}$ production, which may provide a good opportunity to search for $\Upsilon\to \eta_c+\gamma$
in experiment. In addition, the decay width for $\chi_{c1}$ production is several times larger than those for $\chi_{c0,2}$.
Finally, we find the NLO NRQCD prediction for the branching fraction of $\Upsilon\to \chi_{c1}+\gamma$ is
only half of the lower bound of the experimental data measured recently by {\tt Belle}.
Moreover, there exists serious contradiction between theory and experiment for $\Upsilon\to \eta_c+\gamma$.
The discrepancies between theory and experiment deserve further research efforts.
}


\begin{document}
\maketitle
\section{Introduction}

Heavy quarkonium, composed of a pair of heavy quark and heavy antiquark,
is the coexistence of several distinct mass scales, which renders it
a clean platform to probe the interplay between perturbative and nonperturbative dynamics.
The process of a bottomonium radiative decay into a charmonium involves two flavors of heavy quark, which
may enrich our knowledge on quarkonium structure and the nonperturbative properties of QCD.

Many efforts have been devoted to searching for the hint of the process in experiment.
The search for charmonium and charmonium-like states through $\Upsilon$ radiative decay has
been performed by {\tt Belle} collaboration~\cite{Belle:2010sgx,Belle:2011jbs,Shen:2012rn}. Recently,
the first observation of the radiative decay of the $\Upsilon$ into a $P$-wave charmonium has been reported with $6.3$ standard deviations  by {\tt Belle} collaboration~\cite{Belle:2019ybg}:
${\rm Br}(\Upsilon\to \chi_{c1}+\gamma)=4.7^{+2.4}_{-1.8}({\rm stat})^{+0.4}_{-0.5}({\rm sys})\times 10^{-5}$.

On the theory side, it is believed that a bottomonium radiative decay into a charmonium can
be systematically described by the nonrelativistic QCD (NRQCD) factorization approach~\cite{Bodwin:1994jh}, which is based on the
effective-field-theory formalism and directly linked with the first principles of QCD.
Based on this well-established framework, the production rates of $\eta_c$ and
$\chi_{cJ}$ with $J=0,1,2$ in the $\Upsilon$ radiative decays have been calculated at leading order (LO)
 in the strong coupling constant $\alpha_s$ in Ref.~\cite{Gao:2007fv}. Very recently, the processes have been restudied by the authors
 in Ref.~\cite{Shen:2020woq}, where the analytical expressions were presented in terms of the one-loop scalar Passarino-Veltman integrals. Similarly,
the radiative decays of $\eta_b$ or $\chi_{bJ}$ into $J/\psi$
have also been computed in Refs.~\cite{Gao:2007fv,Hao:2006nf}.

It is quite disquieting that the theoretical prediction on the cross section of $\Upsilon\to \chi_{c1}+\gamma$ is much
smaller than the experimental data~\cite{Belle:2019ybg}.
Even worse, the LO theoretical prediction bears a strong renormalization scale dependence.

Recently, there have been great advances in the calculation of high-order radiative corrections
to quarkonium inclusive and exclusive production and decay~\cite{Czarnecki:1997vz,Beneke:1997jm,Marquard:2014pea,Beneke:2014qea,Czarnecki:2001zc,Feng:2015uha,Onishchenko:2003ui,
Chen:2015csa,Sang:2015uxg,Feng:2017hlu,Wang:2018lry,Chen:2017pyi,Feng:2019zmt,Yu:2020tri,Sang:2020fql,Yang:2020pyh}.
For many processes, the two-loop radiative corrections are significant relative to the LO results,
which may substantially change the theoretical
predictions~\cite{Czarnecki:1997vz,Beneke:1997jm,Marquard:2014pea,Beneke:2014qea,Czarnecki:2001zc,Feng:2015uha,Feng:2017hlu}.
In contrast, the
next-to-next-to-leading-order (NNLO) corrections to the cross sections of
$J/\psi+\eta_c$ and $\eta_c(\chi_{cJ})+\gamma$ production at B factory are moderate~\cite{Feng:2019zmt,Chen:2017pyi,Sang:2020fql}.
More importantly, by incorporating the NNLO corrections, the NRQCD theoretical predictions for $e^+e^-\to J/\psi+\eta_c$ and
$e^+e^-\to \chi_{c1}+\gamma$ can be consistent with the experimental measurements.
Encouraged by the success of the NRQCD predictions on these two processes, we compute the next-to-leading-order (NLO) corrections for $\Upsilon\to \eta_c(\chi_{cJ})+\gamma$ in this work.
For the process of a vector bottomonium radiative decay into a charmonium, the bottomonium can decay
through either a photon or two gluons associated with a photon, which are called the QED process and the QCD process
respectively.
Since the Feynman diagrams for the QCD process start at one loop, we need to compute the two-loop corrections.
For the QED process, we only need to include the one-loop corrections.

The remainder of this paper is organized as follows.
In section~\ref{sec-general-formula}, we present the general formulas for the helicity amplitudes and partial widths of
$\Upsilon\to \eta_c(\chi_{cJ})+\gamma$. In section~\ref{sec-nrqcd},
after outlining the NRQCD factorization formula, we describe the theoretical strategy to derive
the NRQCD short-distance coefficients (SDCs).
In section~\ref{sec-calculation}, we describe the technicalities encountered in the calculation.
To reduce the theoretical uncertainty,
we introduce two schemes to compute the QED process in section~\ref{sec-two-schemes}.
Our main numerical results for the SDCs of various helicity amplitudes are presented in
section~\ref{sec-main-results}.
Section~\ref{sec-phen} is devoted to the phenomenological analysis and discussion.
Finally, we summarize in section~\ref{sec-summary}.
In Appendix~\ref{appendix-helicity-projectors}, we present the explicit expressions of various helicity
projectors introduced in section~\ref{sec-nrqcd}. In Appendix~\ref{appendix-Cheng-Wu}, we describe the
approach of dealing with loop integrals by Cheng-Wu theorem.

\section{The general formulas\label{sec-general-formula}}

It is of some advantage to utilize the helicity amplitude formalism to analyze the
hard exclusive decay process.
In general, we can express the helicity partial width of $\Upsilon\to H+\gamma$ ($H$ can be $\eta_c$ or $\chi_{cJ}$) as
\beq\label{eq-gen-rate-helicity}
\Gamma(\Upsilon\to H(\lambda_1)+\gamma(\lambda_2))=\frac{1}{3}\frac{1}{2m_\Upsilon}\frac{1}{8\pi}\frac{2|{\bf P}|}{m_\Upsilon}|A^{(H)}_{\lambda_1,\lambda_2}|^2,
\eeq
where $m_{\Upsilon}$ represents the mass of $\Upsilon$ meson and $P$ denotes
the momentum of the charmonium $H$. The magnitude of the three momentum of $P$ is
determined by
\beq
|{\bf P}|=\frac{\lambda^{1/2}(m_\Upsilon^2,m_H^2,0)}{2m_\Upsilon}=\frac{m_\Upsilon^2-m_H^2}{2m_\Upsilon},
\eeq
where the K\"allen function is defined via $\lambda(x,y,z)=x^2+y^2+z^2-2xy-2xz-2yz$. The spin average for the
initial meson has been done, and the integration over the two-body phase space is explicitly carried out.

In eq.~(\ref{eq-gen-rate-helicity}), $\lambda_1$ and $\lambda_2$ signify the helicities of the charmonia $H$ and the outgoing photon. According to the angular momentum conservation, we can enumerate all the helicity amplitudes $A^{(\eta_c)}_{0,\pm 1}$, $A^{(\chi_{c0})}_{0,\pm 1}$, $A^{(\chi_{c1})}_{\pm1,\pm 1}$, $A^{(\chi_{c1})}_{0,\pm 1}$, $A^{(\chi_{c2})}_{\pm2,\pm 1}$, $A^{(\chi_{c2})}_{\pm1,\pm 1}$ and $A^{(\chi_{c2})}_{0,\pm 1}$, which, however, are not independent. We
can reduce the number of the helicity amplitudes by applying the relations of the parity invariance~\cite{Haber:1994pe} $A^{(\eta_c)}_{\lambda_1,\lambda_2}=-A^{(\eta_c)}_{-\lambda_1,-\lambda_2}$, and $A^{(\chi_{cJ})}_{\lambda_1,\lambda_2}=(-1)^J A^{(\chi_{cJ})}_{-\lambda_1,-\lambda_2}$. Explicitly, we have
\begin{subequations}\label{eq-helicity-parity-invariance}
\bqa
A^{(\eta_c)}_{0,1}&=&-A^{(\eta_c)}_{0,-1}, \\
A^{(\chi_{c0})}_{0,1}&=&A^{(\chi_{c0})}_{0,-1},  \\
A^{(\chi_{c1})}_{0,1}&=&-A^{(\chi_{c1})}_{0,-1}, \quad A^{(\chi_{c1})}_{1,1}=-A^{(\chi_{c1})}_{-1,-1}, \\
A^{(\chi_{c2})}_{0,1}&=&A^{(\chi_{c2})}_{0,-1}, \quad A^{(\chi_{c2})}_{1,1}=A^{(\chi_{c2})}_{-1,-1}, \quad A^{(\chi_{c2})}_{2,1}=A^{(\chi_{c2})}_{-2,-1}.
\eqa
\end{subequations}

In the limit of $m_b\gg m_c$, the helicity amplitudes in (\ref{eq-helicity-parity-invariance}) satisfy the asymptotic behavior
\bqa\label{eq-helicity-selection-rule}
A^{(H)}_{\lambda_1,\lambda_2}\propto r^{1+|\lambda_1|},
\eqa
where $r$ is defined via $r=m_c/m_b$. One power of $r$ in eq.~(\ref{eq-helicity-selection-rule}) originates from the large
momentum transfer which is required for the heavy-quark pair to form the heavy quarkonium with small relative
momentum and the other powers arise from the helicity selection rule in perturbative QCD~\cite{Chernyak:1980dj,Brodsky:1981kj}.

In terms of the independent helicity amplitudes, the total partial widths can be expressed as
\begin{subequations}\label{eq-gen-rate-helicity-explicit}
\bqa
\Gamma(\Upsilon\to \eta_c+\gamma)&=&\frac{1}{3}\frac{1}{2m_\Upsilon}\frac{1}{8\pi}\frac{2|{\bf P}|}{m_\Upsilon}2\times |A^{(\eta_c)}_{0,1}|^2,\\
\Gamma(\Upsilon\to \chi_{c0}+\gamma)&=&\frac{1}{3}\frac{1}{2m_\Upsilon}\frac{1}{8\pi}\frac{2|{\bf P}|}{m_\Upsilon}2\times |A^{(\chi_{c0})}_{0,1}|^2,\\
\Gamma(\Upsilon\to \chi_{c1}+\gamma)&=&\frac{1}{3}\frac{1}{2m_\Upsilon}\frac{1}{8\pi}\frac{2|{\bf P}|}{m_\Upsilon}2\times \bigg(|A^{(\chi_{c1})}_{0,1}|^2+|A^{(\chi_{c1})}_{1,1}|^2\bigg),\\
\Gamma(\Upsilon\to \chi_{c2}+\gamma)&=&\frac{1}{3}\frac{1}{2m_\Upsilon}\frac{1}{8\pi}\frac{2|{\bf P}|}{m_\Upsilon}2\times \bigg(|A^{(\chi_{c2})}_{0,1}|^2+|A^{(\chi_{c2})}_{1,1}|^2+|A^{(\chi_{c2})}_{2,1}|^2\bigg).
\eqa
\end{subequations}

\section{NRQCD formalism for $\Upsilon\to H+\gamma$~\label{sec-nrqcd}}


According to the NRQCD formalism~\cite{Bodwin:1994jh}, we can factorize the helicity amplitude into
\bqa\label{eq-nrqcd}
A^{(H)}_{\lambda_1,\lambda_2}=\sqrt{2m_\Upsilon}\sqrt{2m_H}c_{\lambda_1,\lambda_2}(H)
\frac{\sqrt{\langle \mathcal{O}_{\Upsilon} \rangle_\Upsilon}}{m_b^{3/2}} \frac{\sqrt{\langle\mathcal{O}_{H}\rangle_H}}{m_c^n},
\eqa
where $n=\tfrac{3}{2}$ for $\eta_c$ and $n=\tfrac{5}{2}$ for $\chi_{cJ}$, $c_{\lambda_1,\lambda_2}(H)$ signifies the dimensionless SDC of the corresponding helicity amplitude, and the NRQCD long-distance matrix elements (LDMEs) are defined via
$\langle \mathcal{O}_{h} \rangle_h\equiv \langle h|\psi^{\dagger}{\mathcal K}_h \chi \chi^\dagger {\mathcal K}_h \psi |h\rangle$ with
\begin{subequations}
\begin{eqnarray}
{\mathcal K}_{\Upsilon}&=&{\bm \sigma},\\
{\mathcal K}_{\chi_{c0}}&=&\frac{1}{\sqrt{3}}\left(-\frac{i}{2}\overleftrightarrow{{\bf
D}}\cdot {\bm\sigma}\right),\\
{\mathcal K}_{\chi_{c1}}&=&\frac{1}{\sqrt{2}}\left(-\frac{i}{2}\overleftrightarrow{{\bm
D}}\times{\bm \sigma}\right),\\
{\mathcal K}_{\chi_{c2}}&=&-{i\over 2}\overleftrightarrow{D}^{(i}\sigma^{j)}.
\end{eqnarray}
\end{subequations}
In $\langle \mathcal{O}_{h} \rangle_h$, $\psi$  and $\chi$ are the Pauli spinor fields annihilating a heavy quark and antiquark respectively. In eq.~(\ref{eq-nrqcd}),
we have made used of the vacuum-saturation approximation to relate the LDMEs for charmonium decay to those for production~\cite{Bodwin:1994jh}
\bqa\label{eq-vacuum-approximation}
\langle\mathcal{O}_{H}\rangle_H=\langle 0| \chi^{\dagger}{\mathcal K}_H \psi | H \rangle\langle H | \psi^\dagger {\mathcal K}_H \chi |0 \rangle \bigg(1+\mathcal{O}(v^4)\bigg),
\eqa
where $v$ denotes the typical velocity of the charm quark in charmonium.

In eq.~(\ref{eq-nrqcd}), the factor $\sqrt{2m_\Upsilon}\sqrt{2m_H}$ appears on the right side because
we adopt relativistic normalization for the meson states $\Upsilon$ and $H$, but we use conventional
nonrelativistic normalization for the NRQCD LDMEs. Since we do not consider the relativistic corrections in current work,
it is reasonable to take the approximation $m_{\Upsilon}\approx 2m_b$ and $m_H\approx 2m_c$.
In addition, it is straightforward to deduce the helicity selection rule for SDCs
\bqa\label{eq-helicity-scaling-rule-1}
c_{\lambda_1,\lambda_2}(H)\propto r^{1/2+|\lambda_1|}
\eqa
from (\ref{eq-helicity-selection-rule}) by noticing that $\sqrt{\langle\mathcal{O}_{H}\rangle_H}\propto m_c^n$.

Since the NRQCD SDCs are insensitive to the nonperturbative hadronization effects, they
can be determined with the aid of the standard perturbative matching technique. That is, by
replacing the physical $\Upsilon$ with a fictitious onium $b\bar{b}({^3S_1})$, and replacing $H$ meson with $\widetilde{H}$ composed of a
free $c\bar{c}$ pair and carrying the same quantum numbers as $H$, {\emph i.e}., $\widetilde{H}=c\bar{c}(^1S_0)$ for $\eta_c$ and $\widetilde{H}=c\bar{c}(^3P_J)$ for $\chi_{cJ}$.
Explicitly, we have
\bqa\label{eq-nrqcd-per}
{\mathcal A}^{(\widetilde{H})}_{\lambda_1,\lambda_2}&=&c_{\lambda_1,\lambda_2}(H)
\frac{\sqrt{\langle b\bar{b}(^3S_1)|\mathcal{O}_{\Upsilon}|b\bar{b}(^3S_1)\rangle}}{m_b^{3/2}} \frac{\sqrt{\langle \widetilde{H}|\mathcal{O}_{H}|\widetilde{H}\rangle}}{m_c^n}\nn\\
&=&c_{\lambda_1,\lambda_2}(H)2N_c\frac{2m_c2m_b}{m_b^{3/2}m_c^n},
\eqa
 where we use the relativistic normalization for the heavy quark states in the computation of full QCD amplitude ${\mathcal A}^{(\widetilde{H})}_{\lambda_1,\lambda_2}$ and NRQCD matrix elements. An additional
factor $2N_c$ arises from the spin and color factors of the NRQCD matrix elements.

 From eq.~(\ref{eq-nrqcd-per}), we are able to reexpress the SDC as
\bqa\label{eq-sdcs}
c_{\lambda_1,\lambda_2}(H)={\mathcal A}^{(\widetilde{H})}_{\lambda_1,\lambda_2}\frac{m_b^{1/2}m_c^{n-1}}{8N_c}.
\eqa

To evaluate the QCD amplitude ${\mathcal A}^{(\widetilde{H})}_{\lambda_1,\lambda_2}$,
we assign the momenta
of the $b$ and $\bar{b}$ quarks to be $\frac{Q}{2}$ and the momenta of the $c$ and $\bar{c}$ quarks to be
 \begin{subequations}\label{eq-kinematics-momenta}
 \bqa
p&=&\frac{P}{2}+q, \\
\bar{p}&=&\frac{P}{2}-q,
\eqa
\end{subequations}
where $P$ and $q$ denote the total momentum of the $c\bar{c}$ pair and the relative momentum, respectively.
The on-shell condition $p^2=\bar{p}^2=m_c^2$ enforces that
\bqa
\label{kinematics-on-shell}
&&P\cdot q=0,\qquad P^2=4m_c^2.
\eqa

It is convenient to employ the covariant spin-projector and color-projector to enforce the $b\bar{b}$ and $c\bar{c}$ pairs in
the spin-singlet/spin-triplet and color-singlet states~\cite{Petrelli:1997ge,Bodwin:2013zu}.
The relativistically normalized color-singlet/spin-triplet projector for $\Upsilon$ meson reads
\bqa
\label{spin-projector-bb}
\overline{\Pi}_1=\frac{-1}{\sqrt{2}}(\frac{Q\!\!\!\!\slash}{2}+m_b){\epsilon}_\Upsilon\!\!\!\!\!\!\slash\,\,\otimes {{\bf 1}_c \over \sqrt{N_c}},
\eqa
where $\epsilon_{\Upsilon}$ denotes the polarization vector of $\Upsilon$.
The relativistically normalized color-singlet and spin-singlet/spin-triplet projectors for charmonium read~\cite{Bodwin:2013zu}
 \begin{subequations}\label{spin-projector-cc}
\bqa
\Pi_0&=&\frac{1}{8\sqrt{2}m_c^2}(\pbarslash-m_c)\gamma^5(\Pslash+2m_c)(\pslash+m_c)\otimes {{\bf 1}_c \over \sqrt{N_c}},\\
\Pi_1^\mu&=&\frac{-1}{8\sqrt{2}m_c^2}(\pbarslash-m_c)\gamma^\mu(\Pslash+2m_c)(\pslash+m_c)\otimes {{\bf 1}_c \over \sqrt{N_c}}.
\eqa
\end{subequations}

The amplitude of $b\bar{b}(^3S_1)\to c\bar{c}(^1S_0)+\gamma$ can be projected out by directly replacing the bottom spinors $u(Q/2)\bar{v}(Q/2)$ with $\overline{\Pi}_1$ and replacing the charm spinors $v(\bar{p})\bar{u}(p)$ with $\Pi_0$, namely
 \bqa
\label{s-wave-projector}
\mathcal{A}^{(c\bar{c}(^1S_0))}={\rm Tr}[\overline{\Pi}_1 \mathcal{A}\Pi_0]\Big |_{q=0},
\eqa
where $\mathcal{A}$ represents the quark-level amplitude of $b\bar{b}\to c\bar{c}+\gamma$ with the external quark spinors truncated.
Since we are only concerned with the SDCs at the lowest order in velocity expansion, the relative momentum $q$ is set to zero.

The amplitude of $b\bar{b}(^3S_1)\to c\bar{c}({}^3P_J)+\gamma$ can be projected out by differentiating
the color-singlet/spin-triplet quark-level amplitude with respect to the relative momentum $q$,
followed by setting $q$ to zero:
 \bqa
\label{p-wave-projector}
\mathcal{A}^{(c\bar{c}({}^3P_J))}=\epsilon^{*(J)}_{\mu\nu}\frac{d}{dq_\nu}{\rm Tr}[\overline{\Pi}_1 \mathcal{A} \Pi_1^\mu]\Big |_{q=0},
\eqa
with $\epsilon^{(J)}_{\mu\nu}$ denoting the polarization vectors affiliated with $J=0,1,2$.

In order to further extract the helicity amplitudes, it is very useful to construct various helicity projectors ${\mathcal P}_{\lambda_1,\lambda_2}^{(H)}$, the explicit expressions of which can be found in Appendix~\ref{appendix-helicity-projectors}.
Making use of the helicity projectors, we can readily obtain the helicity amplitudes ${\mathcal A}^{(\widetilde{H})}_{\lambda_1,\lambda_2}$ through first replacing the polarization vectors of $\Upsilon$ and $H$ with the corresponding ${\mathcal P}_{\lambda_1,\lambda_2}^{(H)}$, and then contracting the Lorentz indices.

Now we have collected all the necessary ingredients to evaluate the quark-level helicity amplitudes ${\mathcal A}^{(\widetilde{H})}_{\lambda_1,\lambda_2}$ in perturbative QCD.
It is then straightforward to ascertain the SDCs following eq.~(\ref{eq-sdcs}), and further compute the physical helicity amplitudes
 (\ref{eq-nrqcd}) and partial widths (\ref{eq-gen-rate-helicity}).

\section{Technical strategy of calculating SDCS \label{sec-calculation}}

In this section, we describe the computational technicalities utilized in evaluating the QCD amplitude and
the SDCs.

\begin{figure}[htbp]
 	\centering
 \includegraphics[width=0.7\textwidth]{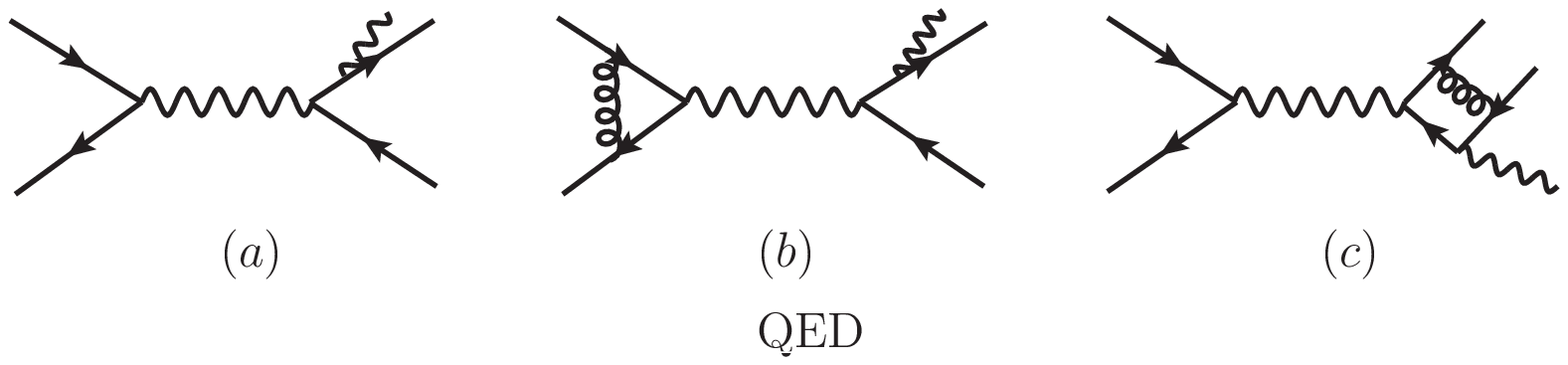}
 \includegraphics[width=0.8\textwidth]{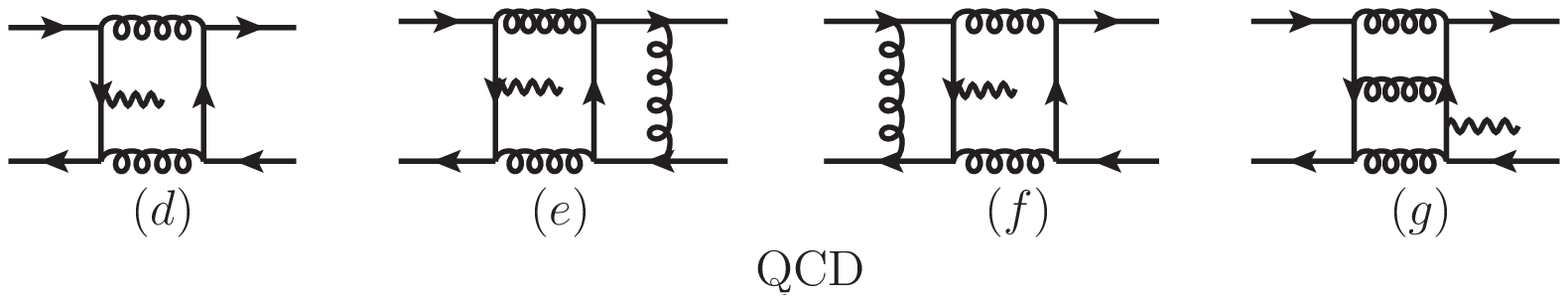}
 \caption{Some representative Feynman diagrams for $\Upsilon\to \eta_c(\chi_{cJ})+\gamma$. ($a$) corresponds to the LO QED contribution,
 ($b$) and ($c$) correspond to the NLO QED contribution, ($d$) corresponds to the LO QCD contribution, and ($e$)-($g$) correspond to the NLO QCD contribution. The Feynman diagrams are drawn by \texttt{JaxoDraw}~\cite{Binosi:2008ig}.
 \label{fig-feynman-diagram}}
 \end{figure}

We use {\tt FeynArts}~\cite{Hahn:2000kx} to generate the Feynman diagrams for $b\bar{b}\to c\bar{c}+\gamma$ and corresponding Feynman amplitudes through NLO in $\alpha_s$.  Some typical QED and QCD Feynman diagrams are illustrated in figure~\ref{fig-feynman-diagram}.
Due to the color-singlet constraint and C-parity conservation, the QCD Feynman diagrams
start at one-loop order. In this work, we will compute both contributions through relative order $\mathcal{O}(\alpha_s)$, {\emph i.e}.,
calculating the QED contribution through one-loop order and QCD contribution through two-loop order.
Employing the color-singlet/spin-triplet projectors (\ref{spin-projector-bb}) and (\ref{spin-projector-cc}),
following the recipe as depicted in section~\ref{sec-nrqcd}
to single out the $S$ or $P$-wave components of the amplitude, we are able to project out the intended $b\bar{b}(^3S_1)\to c\bar{c}
(^1S_0/ ^3P_J)+\gamma$ amplitude, order by order in $\alpha_s$.
We then employ the packages {\tt FeynCalc}~\cite{Mertig:1990an,Shtabovenko:2016sxi} and {\tt FormLink}~\cite{Feng:2012tk} to carry out various helicity amplitudes with the aid of the helicity projectors specified in Appendix~\ref{appendix-helicity-projectors}.

It is quite straightforward to compute the LO QED SDCs, which will be analytically presented in section~\ref{sec-main-results}.
To determine the SDCs at loop level, we adopt the standard approach to directly extract the SDCs,  {\emph i.e}.,  we implement the
procedures (\ref{s-wave-projector}) for $\eta_c$ and (\ref{p-wave-projector}) for $\chi_{cJ}$
prior to conducting the loop integral, which amounts to directly extracting the contribution from the hard region
in the context of strategy of region~\cite{Beneke:1997zp}.
We apply the dimensional regularization to regularize both UV and IR divergences throughout this work.
For the loop integrals, we utilize the packages {\tt Apart}~\cite{Feng:2012iq} and {\tt FIRE}~\cite{Smirnov:2014hma} to conduct partial fraction and the corresponding integration-by-part (IBP) reduction. We end up with 13 one-loop master integrals (MIs)
and 567 two-loop MIs, of which 237 are complex-valued integrals.
For the one-loop MIs, one can work out the analytical expressions with the aid of {\tt Package-X}~\cite{Patel:2015tea}.

It becomes rather challenging to deduce the analytical expressions for all the encountered two-loop MIs.
Instead, we are content with high-precision numerical results which actually are enough for the phenomenological analysis.
For the two-loop MIs, we make use of the {\tt FIESTA}~\cite{Smirnov:2013eza} to perform sector decomposition (SD), which are initially introduced in Refs.~\cite{Binoth:2000ps,Binoth:2003ak,Binoth:2004jv,Heinrich:2008si,Bogner:2007cr,Bogner:2008ry}.
For the real-valued MIs, we directly use {\tt CubPack}~\cite{CubPack} to carry out the numerical integration.
In contrast to the application of SD to the Euclidean region, the singularities encountered in the physical region lie
inside, rather than sit on, the integration boundary, which render the integrals hard to be numerically evaluated.
We can overcome the difficulty by deforming the integration contour through the following variable
transformation~\cite{Borowka:2012qfa}:
\begin{eqnarray}
\label{contour-deformation}
z_k=x_k-i\lambda_k x_k (1-x_k)\frac{\partial F}{\partial x_k},
\end{eqnarray}
where $F$ denotes the $F$-term in the $\alpha$ parametrization, $\lambda_k$ is some positive number. Actually, the integration efficiency may vary drastically with ${\lambda_k}$. In our calculation, we first choose a set of ${\lambda_k}$ and utilize {\tt CubPack}
to conduct the first-round rough numerical integration.
For those integrals with large estimated errors, we adjust the values of ${\lambda_k}$ and perform the integration with a fixed number of sample points. The operation will be repeated until we find a relatively optimized values of ${\lambda_k}$, which render the integration error endurable. With the new determined ${\lambda_k}$, the integral will be evaluated once again with the aid of a parallelized integrator {\tt HCubature}~\cite{HCubature} to reach the desired precision.
For more technical detail, we refer the readers to Ref.~\cite{Feng:2019zmt}. In very rare case, the integrator may generate finite
however incorrect numerical results for some integrals because of the relatively large ${\lambda_k}$, which actually reverses
the sign of the imaginary part of the $F-$term.  This difficulty can be overcome by decreasing the values of ${\lambda_k}$ through
a multiplication factor (less than 1), and reevaluating the integration.

To eliminate the UV divergences at NLO, we perform the renormalization procedure by implementing the $\mathcal{O}(\alpha_s)$ expressions of the renormalization constants $Z_2$ and $Z_m$ accurate up to $\mathcal{O}(\epsilon)$, which can be found in Refs.~\cite{Broadhurst:1991fy,Melnikov:2000zc,Baernreuther:2013caa}.

To guarantee the correctness of the two-loop calculation, we make the following check. We recalculate all the real-valued MIs through another public package {\tt pySecDec}~\cite{Borowka:2017idc}. Meanwhile, we reevaluate most of the complex-valued two-loop MIs (215 of total 237 MIs), which can be
converted into real-valued MIs through the Cheng-Wu theorem~\cite{Cheng-Wu-1,Bjoerkevoll:1992cu,Smirnov_Book}.
The approach of using Cheng-Wu theorem to deal with loop integrals makes the numerical integration much efficient, and
the technical description is presented in Appendix~\ref{appendix-Cheng-Wu}.
The values of the reevaluated MIs are consistent with those computed by {\tt FIESTA} within the errors.
In addition, by using the values of the same sets of MIs, all the divergences in the 7 helicity channels can be removed
by the renormalization constants.

\section{Two schemes to deal with the QED contribution\label{sec-two-schemes}}

As illustrated in figure~\ref{fig-feynman-diagram},
there are two kinds of Feynman diagrams for the processes $\Upsilon\to H+\gamma$:
QED Feynman diagrams represented by figure~\ref{fig-feynman-diagram}($a$)-($c$),
and QCD Feynman diagrams represented by figure~\ref{fig-feynman-diagram}($d$)-($g$).
Correspondingly, we can split the SDCs into two parts
 \bqa
\label{eq-sdcs-qed-qcd}
c_{\lambda_1,\lambda_2}(H)=d_{\lambda_1,\lambda_2}(H)+f_{\lambda_1,\lambda_2}(H),
\eqa
 where $d_{\lambda_1,\lambda_2}(H)$ and $f_{\lambda_1,\lambda_2}(H)$ correspond to the contributions from the QED Feynman
diagrams and the QCD Feynman diagrams respectively.

It is well known, when computing a vector heavy quarkonium decay into a virtual photon, one
confronts very poor convergence in perturbative expansion~\cite{Beneke:1997jm,Marquard:2014pea,Beneke:2014qea}.
Since there exists the similar structure in the QED Feynman diagrams of $\Upsilon\to H+\gamma$, one may worry whether
the higher order perturbative corrections will significantly change the theoretical prediction for the QED contribution.
Actually, we can reduce the theoretical uncertainty by replacing the contribution of $\Upsilon \to \gamma^*$ in the QED
amplitude by the $\Upsilon$ decay constant~\cite{Bodwin:2017pzj}.

The $\Upsilon$ decay constant is defined through the following matrix element of the electro-magnetic current
\begin{eqnarray}
\label{eq-decay-constant-1}
\langle \Upsilon(\epsilon)|\bar{b}\gamma^\mu b|0\rangle=-f_\Upsilon m_\Upsilon \epsilon^{*\mu}_{\Upsilon},
\end{eqnarray}
where the state of $\Upsilon$ meson is relativistically renormalized, and we adopt the same sign convention as in Ref.~\cite{Bodwin:2017pzj}.
The decay constant $f_\Upsilon$ can be related to the $\Upsilon$ electromagnetic decay width through
\begin{eqnarray}
\label{eq-decay-constant-2}
\Gamma(\Upsilon\to e^+e^-)=\frac{4\pi}{3m_\Upsilon}\alpha^2 e_b^2 f_\Upsilon^2,
\end{eqnarray}
where $\alpha$ signifies the electromagnetic coupling constant and $e_b=-1/3$ denotes the charge of bottom quark.
It is straightforward to derive
\begin{eqnarray}
\label{eq-decay-constant-3}
f_\Upsilon = \sqrt{\frac{3m_\Upsilon \Gamma(\Upsilon\to e^+e^-)}{4\pi\alpha^2 e_b^2}}.
\end{eqnarray}

To make distinguishment, we refer to the factorization formula expressed in eq.~(\ref{eq-nrqcd}) as ${\rm QED_{I}}$ scheme,
while the approach by replacing $\Upsilon\to \gamma^*$ with the decay constant $f_\Upsilon$ as ${\rm QED_{II}}$ scheme.
Actually, we can easily convert the ${\rm QED_{I}}$ scheme to the ${\rm QED_{II}}$ scheme by making use of the following substitution
\begin{eqnarray}
\label{eq-decay-constant-2}
\sqrt{\langle \mathcal{O}_\Upsilon \rangle_\Upsilon} \to \sqrt{m_b} f_\Upsilon,
\end{eqnarray}
meanwhile, removing the contributions from the Feynman diagram figure~\ref{fig-feynman-diagram}($b$) in the NLO computation.
For clarity, it is necessary to rewrite the factorization for
the QED helicity amplitude in the ${\rm QED_{II}}$ scheme:
\bqa\label{eq-nrqcd-qed2}
\bigg(A^{(H)}_{\lambda_1,\lambda_2}\bigg)_{\rm QED_{II}}=\sqrt{2m_\Upsilon}\sqrt{2m_H}
d_{\lambda_1,\lambda_2}(H)\frac{f_\Upsilon}{m_b} \frac{\sqrt{\langle\mathcal{O}_{H}\rangle_H}}{m_c^n}.
\eqa
We will make more discussions about the two schemes in section~\ref{sec-main-results} and section~\ref{sec-phen}.

\section{Main results\label{sec-main-results}}

In this section, we present the analytical expressions for the LO QED helicity SDCs and the numerical expressions
for all other SDCs though NLO in $\alpha_s$.
It is convenient to expand the QED and QCD helicity SDCs $d_{\lambda_1,\lambda_2}(H)$ and $f_{\lambda_1,\lambda_2}(H)$ defined in (\ref{eq-sdcs-qed-qcd}) in powers of the strong coupling constant:
\begin{eqnarray}
\label{eq-sdcs-expand-qed}
d_{\lambda_1,\lambda_2}(H)=e^3\bigg[d^{(0)}_{\lambda_1,\lambda_2}(H)+\frac{\alpha_s}{\pi}d^{(1)}_{\lambda_1,\lambda_2}(H)
+\mathcal{O}(\alpha_s^2)\bigg],
\end{eqnarray}
where $e$ denotes the unit of electric charge, and
\begin{eqnarray}
\label{eq-sdcs-expand-qcd}
f_{\lambda_1,\lambda_2}(H)=e\alpha_s^2\bigg[f^{(0)}_{\lambda_1,\lambda_2}(H)+\frac{\alpha_s}{\pi}
\bigg(\frac{1}{2}\beta_0 \ln\frac{\mu_R^2}{m_b^2}\, f^{(0)}_{\lambda_1,\lambda_2}(H)+f^{(1)}_{\lambda_1,\lambda_2}(H)\bigg)
+\mathcal{O}(\alpha_s^2)\bigg],
\end{eqnarray}
where $\beta_0=(11/3)C_A-(4/3)T_Fn_f$ is the one-loop coefficient of the QCD $\beta$ function, with $T_F=\tfrac{1}{2}$ and $n_f$ signifying the number of active quark flavors. In this work, we take $n_f=n_L+n_H$, where $n_L=3$
labels the number of light quark flavors and $n_H=1$ indicates the number of heavy quark flavors. The occurrence of the $\beta_0 \ln \mu_R^2$ in (\ref{eq-sdcs-expand-qcd}) simply reflects the renormalization group invariance.

$d^{(0)}_{\lambda_1,\lambda_2}(H)$ are easy to evaluate and read
 \begin{subequations}\label{eq-sdcs-qed-lo}
\bqa
d^{(0)}_{0,1}(\eta_c)&=&-\frac{r^{1/2}}{27},\\
d^{(0)}_{0,1}(\chi_{c0})&=&\frac{(1-3r^2)r^{1/2}}{27\sqrt{3} (1-r^2)},\\
d^{(0)}_{1,1}(\chi_{c1})&=&\frac{-\sqrt{2}r^{3/2}}{27 (1-r^2)},\;\;\;  d^{(0)}_{0,1}(\chi_{c1})=\frac{\sqrt{2}r^{1/2}}{27 (1-r^2)}, \\
d^{(0)}_{2,1}(\chi_{c2})&=&\frac{-2r^{5/2}}{27 (1-r^2)}, \;\;\; d^{(0)}_{1,1}(\chi_{c2})=\frac{\sqrt{2}r^{3/2}}{27 (1-r^2)},\;\;\;  d^{(0)}_{0,1}(\chi_{c2})=\frac{-\sqrt{6}r^{1/2}}{81 (1-r^2)},
\eqa
\end{subequations}
which manifestly satisfy the helicity selection rule (\ref{eq-helicity-scaling-rule-1}).
It is also not hard to compute $d^{(1)}_{\lambda_1,\lambda_2}(H)$.
Actually, the explicit expressions of $d^{(1)}_{\lambda_1,\lambda_2}(H)$ can be obtained by
combining the results of the $\mathcal{O}(\alpha_s)$ corrections to $\Upsilon\to \gamma^*$ and $\gamma^* \to \eta_c(\chi_{cJ})+\gamma$, the analytical expressions of which can be respectively found in Ref~\cite{Bodwin:1994jh} and
Ref.~\cite{Sang:2009jc}.

It is necessary to discuss the difference between the QED SDCs in the two QED schemes
introduced in section~\ref{sec-two-schemes}. It is obvious that the LO SDCs $d^{(0)}_{\lambda_1,\lambda_2}(H)$ are equal
in the ${\rm QED_{I}}$ and ${\rm QED_{II}}$ schemes. On the other hand,
since all the radiative corrections to $\Upsilon\to \gamma^*$ are encoded in the decay constant,
we must remove the QCD corrections to $\Upsilon\to \gamma^*$  from  $d^{(1)}_{\lambda_1,\lambda_2}(H)$
in the ${\rm QED_{II}}$ scheme.
It is straightforward to derive
\bqa\label{eq-sdcs-qed1-qed2}
\bigg(d^{(1)}_{\lambda_1,\lambda_2}(H)\bigg)_{\rm QED_{II}}=\bigg(d^{(1)}_{\lambda_1,\lambda_2}(H)\bigg)_{\rm QED_{I}}+2C_F d^{(0)}_{\lambda_1,\lambda_2}(H),
\eqa
where the second term on the right side stems from the well-known NLO QCD corrections to $\Upsilon\to \gamma^*$.
In the next section, we will evaluate the partial widths of each helicity in the two schemes and make comparison.

The QCD contribution for $\Upsilon\to H+\gamma$ starts at one-loop order.
The LO SDCs $f^{(0)}_{\lambda_1,\lambda_2}(H)$ can be evaluated analytically with
the aid of the {\tt Package-X}. Very recently, the analytical expressions of the decay widths at LO for the processes $\Upsilon\to H+\gamma$ were presented in terms of the one-loop scalar Passarino-Veltman integrals in Ref.~\cite{Shen:2020woq}.
Since the analytical expressions are rather lengthy and cumbersome, we suppress the explicit forms of $f^{(0)}_{\lambda_1,\lambda_2}(H)$
in this work. Instead, we present the numerical results. As mentioned, it is rather challenging to compute
$f^{(1)}_{\lambda_1,\lambda_2}(H)$ analytically, so we evaluate their values numerically.

To make numerical computation, we choose two sets of typical values of the bottom mass and charm mass, {\emph i.e}., $m_c = 1.483 {\rm GeV},\; m_b=4.580 {\rm GeV}$ and $m_c = 1.68\, {\rm GeV},\;m_b=4.78\, {\rm GeV}$, which correspond to the one-loop and the two-loop heavy quark pole masses respectively~\cite{Feng:2015uha,Bodwin:2017pzj,Sang:2020fql}, converted from the $\overline{\rm MS}$ masses $\overline{m}_c=1.28$ GeV and $\overline{m}_b=4.18$ GeV.

\begin{table}[!htbp]\small
\caption{NRQCD predictions on the dimensionless helicity SDCs for $m_c=1.483$ GeV and $m_b=4.580$ GeV. The QED SDCs are evaluated in the ${\rm QED_{II}}$ scheme. The errors in $f_{\lambda_1,\lambda_2}^{(1)}$ originate from numerical computation.}
\label{tab-sdcs-mc-1.483}
\setlength{\tabcolsep}{0.4mm}
\centering
  \begin{tabular}{|c|c|c|c|c|c|}
   \hline
   \multicolumn{6}{|l|}{$m_{c}=1.483\, {\rm GeV},m_{b}=4.580\, {\rm GeV}$ ;\quad SDCs($\times 10^{-2}$)}\\
   \hline
   \multirow{2}*{$H$} & \multirow{2}*{helicity} & \multicolumn{2}{c|}{QCD}& \multicolumn{2}{c|}{$\rm QED_{II}$}\\
   \cline{3-6}
   &&$f_{\lambda_1,\lambda_2}^{(0)}$&$f_{\lambda_1,\lambda_2}^{(1)}$&$d_{\lambda_1,\lambda_2}^{(0)}$&
   $d_{\lambda_1,\lambda_2}^{(1)}$\\
   \hline
   \multirow{2}*{$\eta_{c}$} & \multirow{2}*{$\lambda_1=0,\lambda_2=1$} &  \multirow{2}*{$-15.14-6.33 i$} &
   $-70.0(3)-116.5(2) i$& \multirow{2}*{$-2.11$}& \multirow{2}*{$3.04+2.80 i$}\\
   &&& $+(6.78+9.02 i)n_L+(5.11+4.76 i)n_H$&& \\
   \hline
   \multirow{2}*{$\chi_{c0}$} & \multirow{2}*{$\lambda_1=0,\lambda_2=1$} &  \multirow{2}*{$-13.56+10.97 i$} &
   $-43.5(8)+62.9(6) i$& \multirow{2}*{$0.93$}&\multirow{2}*{$1.35+1.47 i$}\\
   &&& $+(25.16-3.53 i)n_L+(12.48-8.24 i)n_H$&& \\
   \hline
   \multirow{4}*{$\chi_{c1}$} & \multirow{2}*{$\lambda_1=1,\lambda_2=1$} &  \multirow{2}*{$-4.62$} &
   $-26.6(2)-17.1(2) i$& \multirow{2}*{$-1.08$}&\multirow{2}*{$1.72+1.43 i$}\\
   &&& $+(2.00+0.11 i)n_L+1.13 n_H $&& \\
   \cline{2-6}
    & \multirow{2}*{$\lambda_1=0,\lambda_2=1$} &  \multirow{2}*{$17.89$} &
   $64.4(4)+45.0(4) i$& \multirow{2}*{$3.33$}&\multirow{2}*{$-6.27-3.75 i$}\\
   &&& $+(-11.45-1.26 i)n_L-6.32 n_H$&& \\
   \hline
   \multirow{6}*{$\chi_{c2}$} & \multirow{2}*{$\lambda_1=2,\lambda_2=1$} &  \multirow{2}*{$-0.67+6.11 i$} &
   $33.8(3)-3.2(2) i$& \multirow{2}*{$-0.49$}&\multirow{2}*{$2.31-0.86 i$}\\
   &&& $+(5.55-6.09 i)n_L+(1.80-4.60 i) n_H$&& \\
   \cline{2-6}
    & \multirow{2}*{$\lambda_1=1,\lambda_2=1$} &  \multirow{2}*{$0.43-9.84 i$} &
   $-24.0(3)-8.3(3) i$& \multirow{2}*{$1.08$}&\multirow{2}*{$-5.35+2.67 i$}\\
   &&& $+(-8.00+9.15 i)n_L+(-2.61+7.40 i) n_H$&& \\
     \cline{2-6}
   & \multirow{2}*{$\lambda_1=0,\lambda_2=1$} &  \multirow{2}*{$0.15+15.93 i$} &
   $21.6(5)+14.6(4) i$& \multirow{2}*{$-1.92$}&\multirow{2}*{$8.22-2.44 i$}\\
   &&& $+(11.66-14.64 i)n_L+(3.65-11.98 i) n_H$&& \\
   \hline
  \end{tabular}
 \end{table}

\begin{table}[!htbp]\small
\caption{NRQCD predictions on the dimensionless helicity SDCs for $m_c=1.68$ GeV and $m_b=4.78$ GeV. We use the same conventions as table~\ref{tab-sdcs-mc-1.483}.}
\label{tab-sdcs-mc-1.68}
\setlength{\tabcolsep}{0.4mm}
  \centering
  \begin{tabular}{|c|c|c|c|c|c|}
   \hline
   \multicolumn{6}{|l|}{$m_{c}=1.68\, {\rm GeV},m_{b}=4.78\, {\rm GeV}$ ;\quad SDCs($\times 10^{-2}$)}\\
   \hline
  \multirow{2}*{$H$} & \multirow{2}*{helicity}
   & \multicolumn{2}{c|}{QCD}& \multicolumn{2}{c|}{$\rm QED_{II}$}\\
   \cline{3-6}
   &&$f_{\lambda_1,\lambda_2}^{(0)}$&$f_{\lambda_1,\lambda_2}^{(1)}$&$d_{\lambda_1,\lambda_2}^{(0)}$&
   $d_{\lambda_1,\lambda_2}^{(1)}$\\
   \hline
   \multirow{2}*{$\eta_{c}$} & \multirow{2}*{$\lambda_1=0,\lambda_2=1$} &  \multirow{2}*{$-15.46-7.24 i$} &
   $-65.9(2)-126.5(2) i$&\multirow{2}*{$-2.20$}&\multirow{2}*{$3.38 +2.81 i$}\\
   &&& $+(6.27+9.85 i)n_L+(4.69+5.04 i)n_H$&& \\
   \hline
   \multirow{2}*{$\chi_{c0}$} & \multirow{2}*{$\lambda_1=0,\lambda_2=1$} &  \multirow{2}*{$-13.38+12.54 i$} &
   $-40.6(4)+78.2(3) i$&\multirow{2}*{$0.91$}& \multirow{2}*{$1.62+1.64 i$}\\
   &&& $+(25.35-4.57 i)n_L+(11.95-8.74 i)n_H$&& \\
   \hline
   \multirow{4}*{$\chi_{c1}$} & \multirow{2}*{$\lambda_1=1,\lambda_2=1$} &  \multirow{2}*{$-5.30$} &
   $-27.8(2)-19.0(2) i$& \multirow{2}*{$-1.25$} & \multirow{2}*{$2.16+1.57 i$}\\
   &&& $+(2.45+0.20 i)n_L+1.28 n_H$&& \\
   \cline{2-6}
    & \multirow{2}*{$\lambda_1=0,\lambda_2=1$} &  \multirow{2}*{$18.38$} &
   $63.3(5)+46.3(5) i$& \multirow{2}*{$3.54$} & \multirow{2}*{$-7.09-3.69 i$}\\
   &&& $+(-11.71-1.46 i)n_L-6.08 n_H$&& \\
   \hline
   \multirow{6}*{$\chi_{c2}$} & \multirow{2}*{$\lambda_1=2,\lambda_2=1$} &  \multirow{2}*{$-1.17+6.84 i$} &
   $35.0(2)-4.7(2) i$& \multirow{2}*{$-0.62$} & \multirow{2}*{$2.85-1.10 i$}\\
   &&& $+(6.62-6.22 i)n_L+(2.24-4.77 i) n_H$&& \\
   \cline{2-6}
    & \multirow{2}*{$\lambda_1=1,\lambda_2=1$} &  \multirow{2}*{$1.22-10.53 i$} &
   $-24.1(4)-4.3(4) i$& \multirow{2}*{$1.25$} & \multirow{2}*{$-6.02+3.08 i$}\\
   &&& $+(-9.32+8.97 i)n_L+(-3.21+7.34 i) n_H$&& \\
     \cline{2-6}
   & \multirow{2}*{$\lambda_1=0,\lambda_2=1$} &  \multirow{2}*{$-1.09+15.88 i$} &
   $19.9(5)+5.9(5) i$&\multirow{2}*{$-2.05$}& \multirow{2}*{$8.68-2.79 i$}\\
   &&& $+(12.89-13.40 i)n_L+(4.35-10.07 i) n_H$&& \\
   \hline
  \end{tabular}
 \end{table}
Our main results of the SDCs $d^{(0)}_{\lambda_1,\lambda_2}(H)$, $d^{(1)}_{\lambda_1,\lambda_2}(H)$,
$f^{(0)}_{\lambda_1,\lambda_2}(H)$ and $f^{(1)}_{\lambda_1,\lambda_2}(H)$ are tabulated in
table~\ref{tab-sdcs-mc-1.483} for $m_{c}=1.483\, {\rm GeV},m_{b}=4.580\, {\rm GeV}$ and in table~\ref{tab-sdcs-mc-1.68} for $m_{c}=1.68\, {\rm GeV},m_{b}=4.78\, {\rm GeV}$. In the tables, the QED SDCs are evaluated in the ${\rm QED_{II}}$ scheme. Actually, it is straightforward
to convert the values of $d^{(1)}_{\lambda_1,\lambda_2}$ in the tables into those in the ${\rm QED_{I}}$ scheme by utilizing the formula (\ref{eq-sdcs-qed1-qed2}).
For the sake of reference,
we keep the explicit $n_L$ and $n_H$ dependence in the SDCs.

We have two observations from the tables. First, the LO SDCs $f^{(0)}_{\lambda_1,\lambda_2}(\chi_{c1})$ do not have imaginary parts, which can be well understood by invoking the Cutkosky rule and by noticing the process $\chi_{c1}\to 2 {\rm gluons}$ is
strictly forbidden by the Yang's theorem. Similarly, the imaginary parts of the terms proportional to $n_H$ in $f^{(1)}_{\lambda_1,\lambda_2}(\chi_{c1})$ are also absent, which can be explained by the Yang's theorem and the degenerate phase space for $\chi_{c1}\to c\bar{c}+{\rm gluon}$. Second, we find the LO QCD SDCs of $\chi_{c1}$ roughly satisfy the
helicity selection rule, {\emph i.e}., $|f^{(0)}_{1,1}| \ll |f^{(0)}_{0,1}|$. However, the helicity selection rule is not
so manifest for the counterparts of $\chi_{c2}$, which may be partly due to the fact that the mass of bottom quark is not much larger than that of charm quark.

\section{Phenomenology and discussion\label{sec-phen}}

In this section, we apply the formulas and the numerical results obtained in section~\ref{sec-nrqcd} and section~\ref{sec-main-results}
to make concrete phenomenological analysis.
Prior to making predictions, we need to fix the various input parameters.
We take the QED running coupling constant evaluated at mass of $\Upsilon$ meson, $\alpha(m_\Upsilon)= \tfrac{1}{131}$.
The QCD running coupling constant at different renormalization scales is evaluated with the aid of the package {\tt RunDec}~\cite{Chetyrkin:2000yt}.

We take the masses of the bottomonium and charmonia
\begin{subequations}
\bqa\label{eq-par-mass}
m_{\Upsilon}&=&9.4603\, {\rm GeV},\quad
m_{\eta_c}=2.9839\, {\rm GeV},\\
m_{\chi_{c0}}&=&3.41471\, {\rm GeV},\quad
m_{\chi_{c1}}=3.51067\, {\rm GeV},\quad
m_{\chi_{c2}}=3.55617\, {\rm GeV},
\eqa
\end{subequations}
and the full $\Upsilon$ decay width $\Gamma_{\Upsilon}=54.02\pm 1.25\, {\rm keV}$ from
the latest particle data group (PDG)~\cite{ParticleDataGroup:2020ssz}.
In addition, the $\Upsilon$ decay constant $f_{\Upsilon}=683.8\pm 4.6\, {\rm MeV}$ is determined through
eq.~(\ref{eq-decay-constant-3}).

The NRQCD LDMEs for quarkonia have been obtained through various methods, {\emph i.e}., by fitting experimental data~\cite{Bodwin:2007fz,Guo:2011tz},
by lattice computation~\cite{Bodwin:1993wf,Bodwin:1996tg,Bodwin:2001mk},
and by computation based on the nonrelativistic effective field theories~\cite{Chung:2020zqc}. For more discussion,
we refer the readers to Ref.~\cite{Chung:2020zqc} and the references therein.
In this work, we approximate the NRQCD LDMEs for $S$-wave and $P$-wave quarkonia by the Schr\"odinger
radial wave function at origin and first derivative of the Schr\"odinger
radial wave function at origin respectively,
 \begin{subequations}\label{eq-par-ldme}
\bqa
\langle \mathcal{O}_\Upsilon\rangle_\Upsilon &\approx& \frac{N_c}{2\pi}|R_{1S}^{b\bar{b}}(0)|^2=\frac{N_c}{2\pi}\times 6.477\, {\rm GeV}^3,\\
\langle \mathcal{O}_{\eta_c}\rangle_{\eta_c} &\approx& \frac{N_c}{2\pi}|R_{1S}^{c\bar{c}}(0)|^2=\frac{N_c}{2\pi}\times 0.81\, {\rm GeV}^3,\\
\langle \mathcal{O}_{\chi_{cJ}}\rangle_{\chi_{cJ}} &\approx& \frac{3N_c}{2\pi}|R_{2P}^{\prime c\bar{c}}(0)|^2=\frac{3N_c}{2\pi}\times 0.075\, {\rm GeV}^5,
\eqa
\end{subequations}
where $N_c=3$ signifies the number of color, and we take the values of the radial wave functions from Ref.~\cite{Eichten:1995ch}, which are evaluated based on the Buchm\"uller-Tye (BT) potential model and are frequently used in phenomenological study.

With all these input parameters, we can evaluate the helicity decay widths by eq.~(\ref{eq-gen-rate-helicity}) and
 the total decay widths by eq.~(\ref{eq-gen-rate-helicity-explicit}). The branching fractions can be immediately obtained through dividing the decay widths by the full decay width $\Gamma_{\Upsilon}$.
We tabulate the NRQCD predictions on the helicity partial widths and branching fractions
of $\Upsilon\to H+\gamma$ in table~\ref{tab-decay-rate-mc-1.483} for $m_c=1.483\, {\rm GeV}, \, m_b=4.580\, {\rm GeV}$, and in table~\ref{tab-decay-rate-mc-1.68} for $m_c=1.68\, {\rm GeV}, \, m_b=4.78\, {\rm GeV}$.
To facilitate the comparison,
we use the symbols $\Gamma_{\rm QCD}$ and $\Gamma_{\rm total}$ to represent the pure QCD contribution and the total contribution respectively.
The uncertainties affiliated with the decay widths and the branching fractions are estimated by varying the renormalization scale $\mu_R$ from $2m_c$ to $2m_b$, with the central values evaluated at $\mu_R=m_b$.
In addition, we include the uncertainties from $\Gamma_{\Upsilon}$ and $f_{\Upsilon}$ in the branching fractions.

\begin{table}[htbp]
\caption{NRQCD predictions on the helicity decay widths and branching fractions of $\Upsilon\to H+\gamma$ at LO and NLO in $\alpha_s$ for $m_c=1.483$ GeV and $m_b=4.580$ GeV. The LDMEs are evaluated through the BT potential model.
    $\Gamma_{\rm QCD}$ represents the pure QCD contribution, and $\Gamma_{\rm total}$ denotes the sum of QCD, QED and the interference, where we compute the QED contribution in the ${\rm QED}_{\rm II}$ scheme.
    We include the uncertainties from the renormalization scale in both the decay widths and the branching fractions. In addition,
    we also include the
    uncertainties (the second entry) originating from $\Gamma_{\Upsilon}$ and $f_{\Upsilon}$.}
\label{tab-decay-rate-mc-1.483}
	\setlength{\tabcolsep}{0.3mm}
	\centering
	\begin{tabular}{|c|c|c|c|c|c|c|c|}
		\hline
		\multicolumn{8}{|l|}{$m_{c}=1.483$\,GeV,$m_{b}=4.580$\,GeV ;\quad $\rm\Gamma(\times10^{-1}eV)$}\\
		\hline
		\multicolumn{1}{|c|}{\multirow{2}*{$H$}}
		&\multirow{2}*{helicity}
		&\multicolumn{3}{c|}{LO}
		&\multicolumn{3}{c|}{NLO}
		\\
		\cline{3-8}
		&
		&$\rm\Gamma_{QCD}$
		&$\rm\Gamma_{total}$
		&$\rm Br(\times 10^{-5})$
		&$\rm\Gamma_{QCD}$
		&$\rm\Gamma_{total}$
		&$\rm Br(\times 10^{-5})$
		\\
		\hline
		\multirow{1}*{$\eta_{c}$}
		&$\lambda_{1}=0,\lambda_{2}=1$
		&$12.66^{+8.44}_{-6.05}$
		&$18.91^{+10.05}_{-7.59}$
		&$7.00^{+3.72+0.12}_{-2.81-0.11}$
		&$22.09^{+3.53}_{-5.34}$
		&$28.27^{+3.44}_{-5.70}$
		&$10.47^{+1.27+0.18}_{-2.11-0.17}$
		\\
		\hline
		\multirow{1}*{$\chi_{c0}$}
		&$\lambda_{1}=0,\lambda_{2}=1$
		&$1.74^{+1.16}_{-0.83}$
		&$1.49^{+1.08}_{-0.76}$
		&$0.55^{+0.40+0.009}_{-0.28-0.009}$
		&$1.75^{+0.002}_{-0.17}$
		&$1.57^{+0.01}_{-0.19}$
		&$0.58^{+0.003+0.01}_{-0.07-0.009}$
		\\
		\hline
		\multirow{2}*{$\chi_{c1}$}
		&$\lambda_{1}=1,\lambda_{2}=1$
		&$0.12^{+0.08}_{-0.05}$
		&$0.25^{+0.11}_{-0.09}$
		&\multirow{2}*{$1.31^{+0.62+0.02}_{-0.48-0.01}$}
		&$0.21^{+0.03}_{-0.05}$
		&$0.34^{+0.03}_{-0.06}$
		&\multirow{2}*{$1.44^{+0.02+0.02}_{-0.16-0.02}$}
		\\
		\cline{2-4}	\cline{6-7}
		&$\lambda_{1}=0,\lambda_{2}=1$
		&$1.82^{+1.21}_{-0.87}$
		&$3.28^{+1.57}_{-1.22}$
		&
		&$2.20^{+0.04}_{-0.32}$
		&$3.54^{+0.03}_{-0.37}$
		&
		\\
		\hline
		\multirow{3}*{$\chi_{c2}$}
		&$\lambda_{1}=2,\lambda_{2}=1$
		&$0.21^{+0.14}_{-0.10}$
		&$0.23^{+0.14}_{-0.10}$
		&\multirow{3}*{$0.86^{+0.54+0.01}_{-0.39-0.01}$}
		&$0.15^{+0.02}_{-0.01}$
		&$0.13^{+0.01}_{-0.001}$
		&\multirow{3}*{$0.53^{+0.04+0.009}_{-0.12-0.008}$}
		\\
		\cline{2-4}	\cline{6-7}
		&$\lambda_{1}=1,\lambda_{2}=1$
		&$0.55^{+0.37}_{-0.26}$
		&$0.58^{+0.37}_{-0.27}$
		&
		&$0.42^{+0.01}_{-0.05}$
		&$0.35^{+0.02}_{-0.07}$
		&
		\\
		\cline{2-4}	\cline{6-7}
		&$\lambda_{1}=0,\lambda_{2}=1$
		&$1.44^{+0.96}_{-0.69}$
		&$1.50^{+0.96}_{-0.69}$
		&
		&$1.08^{+0.04}_{-0.20}$
		&$0.96^{+0.07}_{-0.25}$
		&
		\\
		\hline
	\end{tabular}
\end{table}

\begin{table}[h]
\caption{NRQCD predictions on the helicity decay widths and branching fractions of $\Upsilon\to H+\gamma$
at LO and NLO in $\alpha_s$ for $m_c=1.68$ GeV and $m_b=4.78$ GeV. We use the same conventions as in table~\ref{tab-decay-rate-mc-1.68}.}
\label{tab-decay-rate-mc-1.68}
	\setlength{\tabcolsep}{0.5mm}
	\centering
	\begin{tabular}{|c|c|c|c|c|c|c|c|}
		\hline
		\multicolumn{8}{|l|}{$m_{c}=1.68$\,GeV,$m_{b}=4.78$\,GeV ;\quad $\rm\Gamma(\times10^{-1}eV)$}\\
		\hline
		\multicolumn{1}{|c|}{\multirow{2}*{$H$}}
		&\multirow{2}*{helicity}
		&\multicolumn{3}{c|}{LO}
		&\multicolumn{3}{c|}{NLO}
		\\
		\cline{3-8}
		&
		&$\rm\Gamma_{QCD}$
		&$\rm\Gamma_{total}$
		&$\rm Br(\times 10^{-5})$
		&$\rm\Gamma_{QCD}$
		&$\rm\Gamma_{total}$
		&$\rm Br(\times 10^{-5})$
		\\
		\hline
		\multirow{1}*{$\eta_{c}$}
		&$\lambda_{1}=0,\lambda_{2}=1$
		&$9.37^{+4.65}_{-4.44}$
		&$14.16^{+5.59}_{-5.59}$
		&$5.24^{+2.07+0.09}_{-2.07-0.08}$
		&$16.34^{+2.17}_{-3.88}$
		&$20.97^{+2.13}_{-4.15}$
		&$7.76^{+0.79+0.13}_{-1.52-0.12}$
		\\
		\hline
		\multirow{1}*{$\chi_{c0}$}
		&$\lambda_{1}=0,\lambda_{2}=1$
		&$1.03^{+0.51}_{-0.49}$
		&$0.89^{+0.48}_{-0.45}$
		&$0.33^{+0.18+0.006}_{-0.17-0.005}$
		&$1.12^{+0.01}_{-0.136}$
		&$1.03^{+0.02}_{-0.15}$
		&$0.38^{+0.007+0.006}_{-0.05-0.006}$
		\\
		\hline
		\multirow{2}*{$\chi_{c1}$}
		&$\lambda_{1}=1,\lambda_{2}=1$
		&$0.09^{+0.04}_{-0.04}$
		&$0.18^{+0.06}_{-0.06}$
		&\multirow{2}*{$0.78^{+0.27+0.011}_{-0.28-0.011}$}
		&$0.13^{+0.01}_{-0.03}$
		&$0.23^{+0.01}_{-0.04}$
		&\multirow{2}*{$0.84^{+0.01+0.013}_{-0.03-0.012}$}
		\\
		\cline{2-4}	\cline{6-7}
		&$\lambda_{1}=0,\lambda_{2}=1$
		&$1.02^{+0.51}_{-0.48}$
		&$1.93^{+0.68}_{-0.69}$
		&
		&$1.21^{+0.02}_{-0.17}$
		&$2.03^{+0.01}_{-0.19}$
		&
		\\
		\hline
		\multirow{3}*{$\chi_{c2}$}
		&$\lambda_{1}=2,\lambda_{2}=1$
		&$0.15^{+0.07}_{-0.07}$
		&$0.16^{+0.07}_{-0.07}$
		&\multirow{3}*{$0.51^{+0.23+0.008}_{-0.23-0.008}$}
		&$0.10^{+0.002}_{-0.002}$
		&$0.08^{+0.005}_{-0.004}$
		&\multirow{3}*{$0.28^{+0.029+0.005}_{-0.056-0.005}$}
		\\
		\cline{2-4}	\cline{6-7}
		&$\lambda_{1}=1,\lambda_{2}=1$
		&$0.34^{+0.17}_{-0.16}$
		&$0.37^{+0.17}_{-0.17}$
		&
		&$0.24^{+0.01}_{-0.03}$
		&$0.20^{+0.02}_{-0.04}$
		&
		\\
		\cline{2-4}	\cline{6-7}
		&$\lambda_{1}=0,\lambda_{2}=1$
		&$0.76^{+0.38}_{-0.36}$
		&$0.84^{+0.38}_{-0.37}$
		&
		&$0.53^{+0.03}_{-0.09}$
		&$0.48^{+0.05}_{-0.11}$
		&
		\\
		\hline
	\end{tabular}
\end{table}

Comparing table~\ref{tab-decay-rate-mc-1.483} with table~\ref{tab-decay-rate-mc-1.68}, we find the partial widths for
$m_c=1.483\,{\rm GeV}, m_b=4.580\,{\rm GeV}$ are a bit larger than those for $m_c=1.68\,{\rm GeV}, m_b=4.78\,{\rm GeV}$.
It is mainly due to the fact that the helicity amplitude is inversely proportional to the heavy quark masses,
which can be noted in eq.~(\ref{eq-nrqcd}).
From the tables, we notice the $\mathcal{O}(\alpha_s)$ corrections to the
branching fractions are moderate for $\eta_c$ and $\chi_{c2}$, and have minor effects for the $\chi_{c0}$ or $\chi_{c1}$,
which indicates the perturbative expansion in $\alpha_s$ exhibits a decent convergent behavior.
In addition, we note the partial widths are dominated by the QCD contributions for $\chi_{c0}$ and $\chi_{c2}$. In contrast,
the QED part contributes considerably to the partial widths of $\eta_c$ and $\chi_{c1}$, and therefore is indispensable.
Moreover, it is interesting to note that
the total partial width of $\eta_{c}$ is much larger than those of $\chi_{cJ}$, which is mainly caused by the NRQCD LDMEs.
The relatively large branching fraction for the process $\Upsilon\to \eta_c+\gamma$ can provide a good opportunity
for experimental search. Among $\chi_{cJ}$ productions, we find the partial width of $\chi_{c1}$ is several
times larger than those of $\chi_{c0,2}$. Actually, we have the similar observations in $e^+e^-\to \chi_{cJ}+\gamma$,
where the production cross section of $\chi_{c1}$ is much larger than the others~\cite{Sang:2020fql}.
Finally, it is impressive to observe the renormalization scale dependence in the NLO partial widths and
branching fractions are considerably reduced compared to their LO counterparts, especially for $\chi_{cJ}$ production.

Recently, {\tt Belle} collaboration had measured the branching fraction
${\rm Br}(\Upsilon\to \chi_{c1}+\gamma)=4.7^{+2.4+0.4}_{-1.8-0.5}\times 10^{-5}$,
where the two uncertainties correspond to statistical error and systematical error respectively~\cite{Belle:2019ybg}.
It is quite disquieting to find the theoretical prediction with $m_c=1.483$ GeV and $m_b=4.58$ GeV are only half of the lower bound of the experimental data.
Actually, the theoretical predictions are sensitive to the NRQCD LDMEs. Relatively large LDMEs can alleviate the tension
between theory and experiment. In addition, the theoretical predictions depend on the heavy quark masses.
One may increase the theoretical predictions through decreasing the heavy quark masses. However, the
values of the charm and bottom masses may be tuned to be unreliable if we want to explain the experimental measurement.
More seriously, we find the NRQCD prediction on the branching fraction of $\Upsilon\to \eta_c+\gamma$ is several times larger than
the experimental upper limit ${\rm Br}(\Upsilon\to \eta_c+\gamma)< 2.9\times 10^{-5}$~\cite{Belle:2019ybg}.
In addition, we note that the branching
fraction for $\chi_{c1}$ production is larger than the upper limit of $\eta_c$ production. Since the configuration of
the Feynman diagrams for $\chi_{c1}$ is the same as that for $\eta_c$, it is rather unexpected that the production rate for the P-wave $\chi_{c1}$ is larger than that for the S-wave $\eta_c$. There may exist some underlying mechanism which leads to the unexpectation. Actually,
we notice that there is some confliction on ${\rm Br}(\Upsilon\to \chi_{c1}+\gamma)$ between the measurement from Ref.~\cite{Belle:2019ybg} and the measurement from Ref.~\cite{Belle:2010sgx}.
Undoubtedly, the discrepancy between theory and experiment deserves further research efforts.

\begin{figure}[htbp]
 	\centering
 \includegraphics[width=0.45\textwidth]{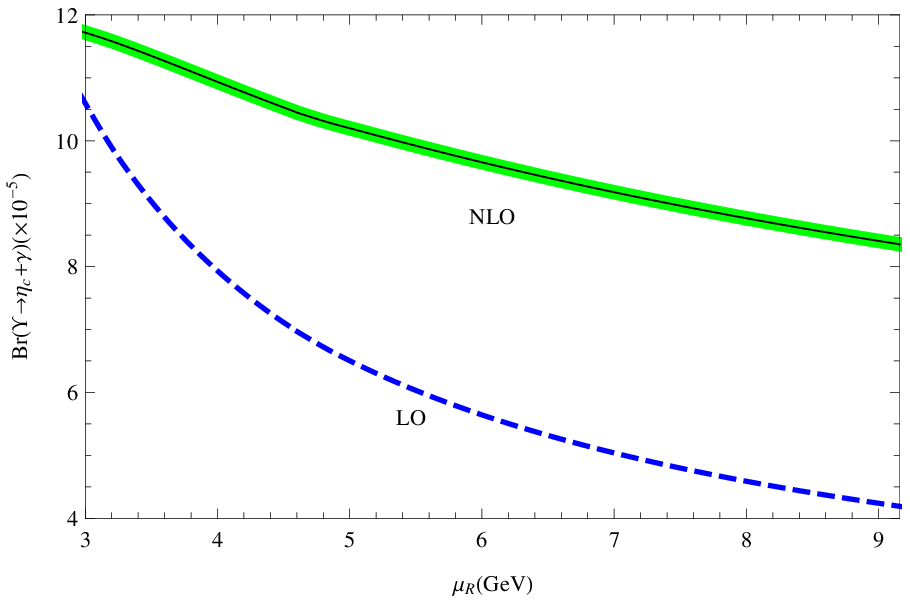}
    \includegraphics[width=0.45\textwidth]{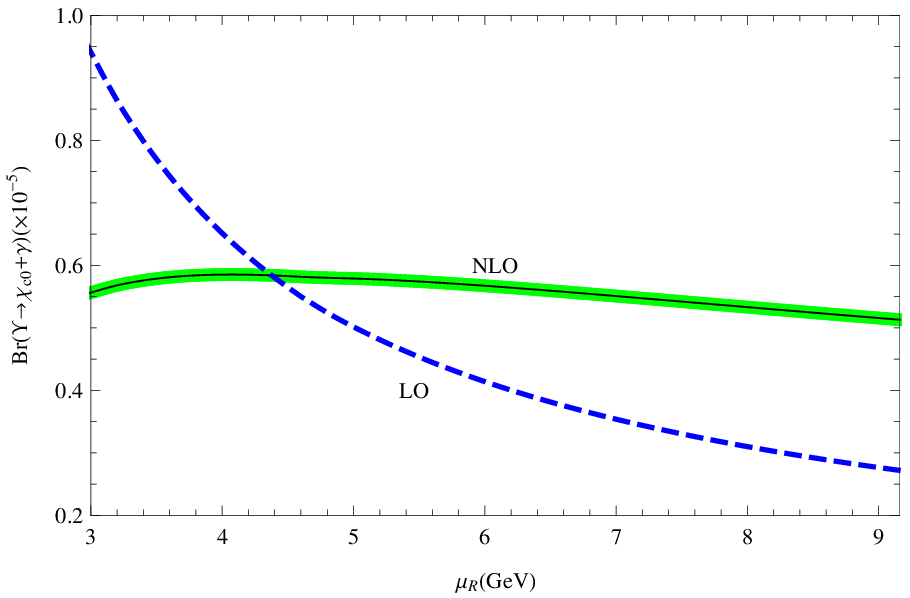}
 \includegraphics[width=0.45\textwidth]{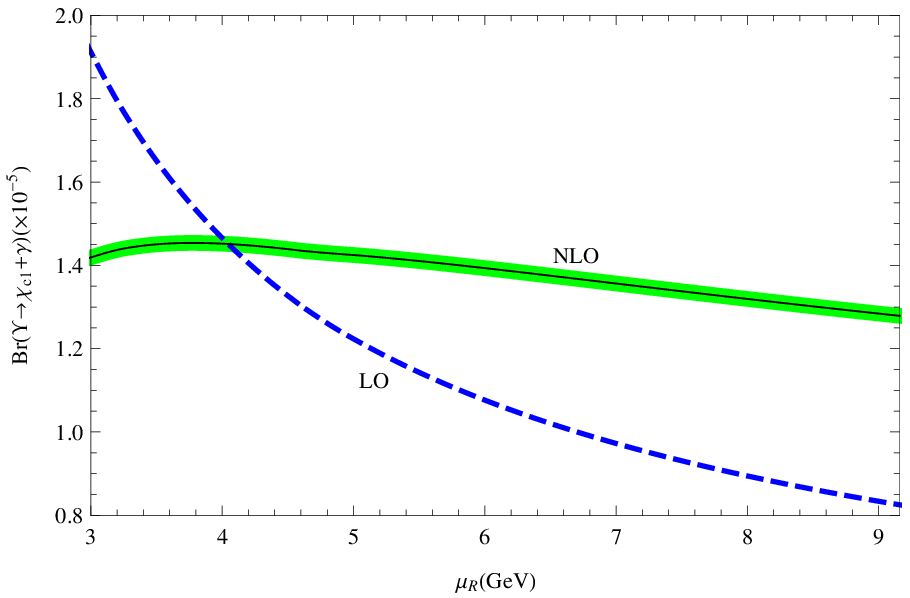}
  \includegraphics[width=0.45\textwidth]{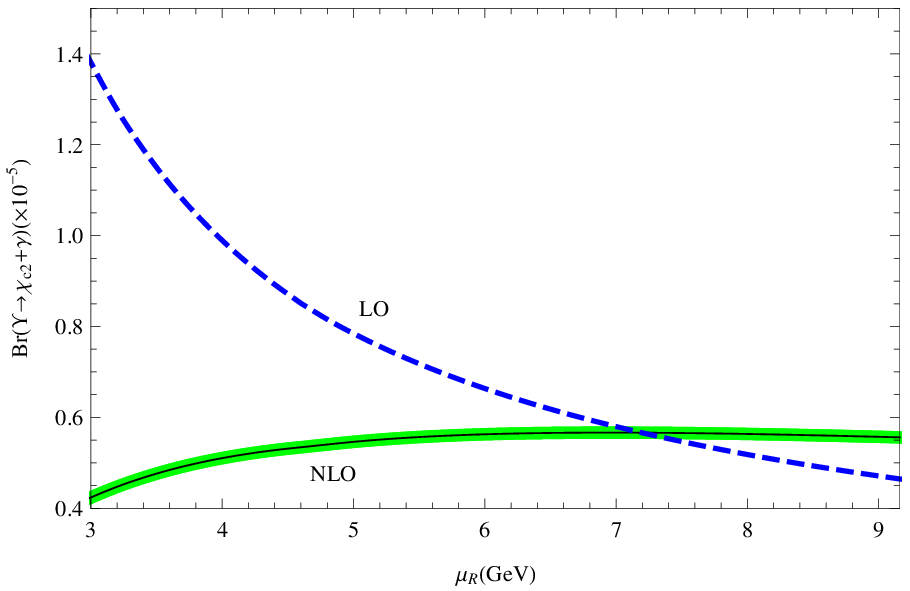}
 \caption{NRQCD predictions for ${\rm Br}(\Upsilon\to H+\gamma)$ as a function of $\mu_R$ accurate up to the LO and NLO in $\alpha_s$.
 The values of the LDMEs are evaluated through the BT potential model. We take $m_c=1.483$ GeV and $m_b=4.58$ GeV. The green band represents the uncertainty band from $\Gamma_{\Upsilon}$ and $f_{\Upsilon}$.
 \label{fig-mu-dependence}}
 \end{figure}

To closely examine the scale dependence, we display the branching fractions of $\Upsilon\to H+\gamma$ as a function
of the renormalization scale $\mu_R$ in figure~\ref{fig-mu-dependence},
where we have taken $m_c=1.483\, {\rm GeV}, \, m_b=4.580\, {\rm GeV}$.
The green band labeled with ``NLO'' corresponds to the uncertainties from $\Gamma_{\Upsilon}$ and $f_{\Upsilon}$, which are actually
rather small.
We find that the NLO prediction
exhibits a very mild $\mu_R$-dependence for $\chi_{c0}$ and $\chi_{c1}$. The feature, to some extent, reflects a decent perturbative convergent behavior.
The NLO prediction can considerably improve the $\mu_R$-dependence for $\chi_{c2}$.
In contrast, we note that the NLO prediction seems to only slightly reduce $\mu_R$-dependence compared to the LO prediction for
$\eta_c$ production, which may be attributed to the sizeable $\alpha_s$ corrections to the QCD contribution of $\eta_c$.

\begin{table}[h]
\caption{Comparison on the decay widths between the two QED schemes. The decay widths include the contributions from QCD, QED, and the interference. In $\Gamma_{\rm QCD+QED_{I}}$, the QED related
contributions are computed in the ${\rm QED_{I}}$ scheme, and in $\Gamma_{\rm QCD+QED_{II}}$, the QED related
contributions are computed in the ${\rm QED_{II}}$ scheme. We estimate the uncertainties
by sliding $\mu_R$ from $2m_c$ to $2m_b$.
\label{tab-comparison-qed1-qed2}}
	\centering
	\begin{tabular}{|c|c|c|c|c|c|}
			\hline
		\multicolumn{6}{|l|}{$m_{c}=1.483$\,GeV,$m_{b}=4.580$\,GeV ;\quad $\rm\Gamma(\times10^{-1}eV)$}\\
		\hline
		\multicolumn{1}{|c|}{\multirow{2}*{$H$}}
		&\multirow{2}*{helicity}
		&\multicolumn{2}{c|}{LO}
		&\multicolumn{2}{c|}{NLO}
		\\
		\cline{3-6}
		&
		&$\rm\Gamma_{QCD+QED_{\Rmnum{1}}}$
		&$\rm\Gamma_{QCD+QED_{\Rmnum{2}}}$
		&$\rm\Gamma_{QCD+QED_{\Rmnum{1}}}$
		&$\rm\Gamma_{QCD+QED_{\Rmnum{2}}}$
		\\
		\hline
		\multirow{1}*{$\eta_{c}$}
		&$\lambda_{1}=0,\lambda_{2}=1$
		&$20.34^{+10.37}_{-7.90}$
		&$18.91^{+10.05}_{-7.59}$
		&$27.96^{+3.16}_{-5.41}$
		&$28.27^{+3.44}_{-5.70}$
		\\
		\hline
		\multirow{1}*{$\chi_{c0}$}
		&$\lambda_{1}=0,\lambda_{2}=1$
		&$1.45^{+1.07}_{-0.74}$
		&$1.49^{+1.08}_{-0.76}$
		&$1.58^{+0.01}_{-0.19}$
		&$1.57^{+0.01}_{-0.19}$
		\\
		\hline
		\multirow{2}*{$\chi_{c1}$}
		&$\lambda_{1}=1,\lambda_{2}=1$
		&$0.28^{+0.12}_{-0.09}$
		&$0.25^{+0.11}_{-0.09}$
		&$0.34^{+0.03}_{-0.05}$
		&$0.34^{+0.03}_{-0.06}$
		\\
		\cline{2-6}
		&$\lambda_{1}=0,\lambda_{2}=1$
		&$3.63^{+1.65}_{-1.29}$
		&$3.28^{+1.57}_{-1.22}$
		&$3.47^{+0.01}_{-0.30}$
		&$3.54^{+0.03}_{-0.37}$
		\\
		\hline
		\multirow{3}*{$\chi_{c2}$}
		&$\lambda_{1}=2,\lambda_{2}=1$
		&$0.23^{+0.15}_{-0.10}$
		&$0.23^{+0.14}_{-0.10}$
		&$0.13^{+0.01}_{-0.003}$
		&$0.13^{+0.01}_{-0.001}$
		\\
		\cline{2-6}
		&$\lambda_{1}=1,\lambda_{2}=1$
		&$0.59^{+0.37}_{-0.27}$
		&$0.58^{+0.37}_{-0.27}$
		&$0.35^{+0.02}_{-0.06}$
		&$0.35^{+0.02}_{-0.07}$
		\\
		\cline{2-6}
		&$\lambda_{1}=0,\lambda_{2}=1$
		&$1.53^{+0.96}_{+0.69}$
		&$1.50^{+0.96}_{-0.69}$
		&$0.95^{+0.06}_{-0.23}$
		&$0.96^{+0.07}_{-0.25}$
		\\
		\hline
	\end{tabular}
\end{table}

In section~\ref{sec-two-schemes}, we have introduced two different schemes to deal with the QED contribution.
It is interesting to compare the decay widths of the two schemes. The results
are illustrated in table~\ref{tab-comparison-qed1-qed2}. We find that the discrepancy
between the predictions of the two schemes is quite insubstantial.
Even for $\eta_c$ and $\chi_{c1}$ production, where the QED contributions are considerable,
the discrepancy is rather negligible.
It might indicate the fact that the NRQCD predictions for the process $\Upsilon\to e^+e^-$ at lower order in $\alpha_s$
can well describe the experimental measurement, even the perturbative expansion is of poor convergence.

Finally, for the sake of completeness, we present the NRQCD predictions for the channels involving radially
excited heavy quarkonia in table~\ref{tab-excited}, which may provide useful help for experimentalists.

\begin{table}[h]
\caption{NRQCD predictions on the branching fraction of a vector bottomonium radiative decay into a charmonium. We
choose $m_c=1.483\, {\rm GeV}, \, m_b=4.580\, {\rm GeV}$ and evaluate the LDMEs through the BT potential model.
The masses of the quarkonia are taken from PDG~\cite{ParticleDataGroup:2020ssz}.
The second row represents the branching fractions up to NLO in $\alpha_s$, where the
uncertainties originate from $\mu_R$-dependence and $\Gamma_{\Upsilon(nS)}$. We display the
experimental data from {\tt Belle}~\cite{Belle:2019ybg} in the third row.
\label{tab-excited}}
	\centering
	\begin{tabular}{|c|c|c|}
		\hline
		Channels&
		Br($\times10^{-5}$)&
		{\tt Belle}-2019~\cite{Belle:2019ybg}
		\\
		\hline
		$\rm \Upsilon\rightarrow\eta_{c}+\gamma$&
		$10.47^{+1.45}_{-2.28}$&
		$<2.9$
		\\
		\hline
		$\rm \Upsilon(2S)\rightarrow\eta_{c}+\gamma$&
		$8.74^{+1.53}_{-2.32}$&
		$-$
		\\
		\hline
		$\rm \Upsilon(3S)\rightarrow\eta_{c}+\gamma$&
		$10.45^{+1.90}_{-2.87}$&
		$-$
		\\
		\hline
		$\rm \Upsilon\rightarrow\eta_{c}(2S)+\gamma$&
		$6.84^{+0.94}_{-1.49}$&
		$<40$
		\\
		\hline
		$\rm \Upsilon\rightarrow\chi_{c 0}+\gamma$&
		$0.58^{+0.01}_{-0.08}$&
		$<6.6$
		\\
		\hline
		$\rm \Upsilon\rightarrow\chi_{c 1}+\gamma$&
		$1.44^{+0.04}_{-0.14}$&
		$4.7^{+2.4+0.4}_{-1.8-0.5}$
		\\
		\hline
		$\rm \Upsilon\rightarrow\chi_{c 2}+\gamma$&
		$0.53^{+0.05}_{-0.11}$&
		$<3.3$
		\\
		\hline
		$\rm \Upsilon(2S)\rightarrow\chi_{c 0}+\gamma$	&
		$0.49^{+0.03}_{-0.04}$&
		$-$
		\\
		\hline
		$\rm \Upsilon(2S)\rightarrow\chi_{c 1}+\gamma$&
		$1.19^{+0.09}_{-0.21}$&
		$-$
		\\
		\hline
		$\rm \Upsilon(2S)\rightarrow\chi_{c 2}+\gamma$&
		$0.45^{+0.05}_{-0.12}$&
		$-$	
		\\
		\hline
		$\rm \Upsilon(3S)\rightarrow\chi_{c 0}+\gamma$&
		$0.60^{+0.04}_{-0.11}$&
		$-$
		\\
		\hline
		$\rm \Upsilon(3S)\rightarrow\chi_{c 1}+\gamma$&
		$1.41^{+0.10}_{-0.25}$&
		$-$
		\\
		\hline
		$\rm \Upsilon(3S)\rightarrow\chi_{c 2}+\gamma$&
		$0.55^{+0.06}_{-0.17}$&
		$-$
		\\
		\hline
	\end{tabular}
\end{table}

\section{Summary~\label{sec-summary}}

In this work, we compute the NLO QCD corrections to the radiative decay $\Upsilon\to \eta_c(\chi_{cJ})+\gamma$
by applying the NRQCD factorization formalism.
We choose two benchmark values of the masses of charm and bottom quarks: $m_c=1.483$ GeV and $m_b=4.580$ GeV, which corresponds to the one-loop heavy quark pole masses, and $m_c=1.68$ GeV and $m_b=4.78$ GeV, which corresponds to the two-loop heavy quark pole masses.
We include both the QED and the QCD Feynman diagrams, the amplitudes of which are calculated up to NLO in $\alpha_s$.
The helicity amplitudes and the helicity decay widths are obtained.
It is the first computation for the processes involving both bottomonium and charmonium at two-loop accuracy.
By utilizing the Cheng-Wu theorem, we can convert most of the complex-valued MIs into real-valued MIs, which
makes the numerical integration much efficient.
In addition, we make use of two different schemes to deal with the QED contribution, and find a good agreement.

Our results indicate that the decay widths are dominated by the QCD contribution for $\chi_{c0}$ and $\chi_{c2}$ production,
while the QED contribution is comparable to the QCD contribution for $\eta_c$ and $\chi_{c1}$ production.
We notice that the $\mathcal{O}(\alpha_s)$ corrections are moderate for $\eta_c$ and $\chi_{c2}$ production,
and are quite marginal for $\chi_{c0}$ and $\chi_{c1}$ production.
It is interesting to note that the NLO corrections considerably reduce the renormalization scale dependence in the branching
fractions of $\chi_{cJ}$ production, which may indicate a decent convergence in perturbative expansion.
With the NRQCD LDMEs evaluated through the BT potential model, we find the decay width of $\eta_c$ production is one-order-of-magnitude larger than those of $\chi_{cJ}$ production, which may provide a good opportunity to search for $\Upsilon\to \eta_c+\gamma$
in experiment. In addition, the decay width of $\chi_{c1}$ is several times larger than those of $\chi_{c0,2}$.
Finally, we find the NRQCD prediction on the branching ratio of $\Upsilon\to \chi_{c1}+\gamma$ is
only half of the lower bound of the experimental data measured by {\tt Belle}.
Moreover, there exists serious contradiction between theory and experiment for $\Upsilon\to \eta_c+\gamma$.
The discrepancies between theory and experiment deserve further research efforts.

\begin{acknowledgments}
The work of Y.-D. Z. and W.-L. S. is supported by the National
Natural Science Foundation of China under Grants No. 11975187 and the Natural Science
Foundation of ChongQing under Grant No. cstc2019jcyj-msxmX0479.
 The work of F. F. is supported by the National Natural
Science Foundation of China under Grant No. 11875318, No. 11505285, and by the Yue
Qi Young Scholar Project in CUMTB.
The work of H.-F. Zhang is supported by the National
Natural Science Foundation of China under Grants No. 11965006.
\end{acknowledgments}

\section*{Appdendix}
\appendix

\section{Construction of various helicity projectors\label{appendix-helicity-projectors}}

In section~\ref{sec-nrqcd}, we have employed the helicity projectors to extract the corresponding helicity
amplitudes of $\Upsilon\to H(\lambda_1)+\gamma(\lambda_2)$. In this appendix, we apply the similar approach
used in Ref.~\cite{Xu:2012uh} to
construct various helicity projectors $\mathcal{P}^{(H)}_{\lambda_1,\lambda_2}$.

For the sake of convenience, we introduce an auxiliary transverse metric tensor and two auxiliary longitudinal vectors,
\begin{subequations}\label{eq-auxiliary}
\begin{eqnarray}
g_{\perp}^{\mu \nu}&=&g^{\mu \nu}+\frac{P^{\mu} P^{\nu}}{|\mathbf{P}|^{2}}-\frac{Q \cdot P}{m_{\Upsilon}^{2}|\mathbf{P}|^{2}}\left(P^{\mu} Q^{\nu}+Q^{ \mu} P^{\nu}\right),\\
L_{\Upsilon}^\mu &=& \frac{1}{|\mathbf{P}|} \bigg(P^\mu-\frac{Q\cdot P}{m_\Upsilon^2}Q^\mu\bigg),\\
L_{\chi_{cJ}}^\mu &=& \frac{1}{|\mathbf{P}|} \bigg(\frac{P\cdot Q}{m_\Upsilon m_{\chi_{cJ}}}P^\mu-\frac{m_{\chi_{cJ}}}{m_\Upsilon}Q^\mu\bigg),
\end{eqnarray}
\end{subequations}
where $P$ and $Q$ denote the momenta of $H$ and $\Upsilon$ mesons respectively, $m_{\chi_{cJ}}$ signifies the mass of $\chi_{cJ}$.
It is obvious the transverse metric tensor satisfies the properties
\begin{subequations}
\begin{eqnarray}
&&g_{\perp \mu \nu} P^{\mu}=g_{\perp \mu \nu} Q^{\mu}=0,\\
&&g_{\perp \mu}^{\mu}=2,\\
&&g_{\perp \mu\alpha}g_\perp^{\alpha\nu}=g_{\perp \mu\alpha}g^{\alpha\nu}=g_{\perp \mu}^\nu.
\end{eqnarray}
\end{subequations}
The longitudinal vectors satisfy $L_{\Upsilon}^\mu Q_\mu=L_{\chi_{cJ}}^\mu P_\mu=0$.

The amplitude $\mathcal{A}^{(\eta_c)}$ of the process $\Upsilon \to \eta_{c}+\gamma$ can be parameterized as
$\mathcal{A}^{(\eta_c)}=\mathcal{A}_{\mu \nu}^{(\eta_c)} \epsilon_{\Upsilon}^{\mu} \epsilon_{\gamma}^{* \nu}(\lambda_{2})$, where $\epsilon_{\Upsilon}$ and $\epsilon_{\gamma}$ denote the polarization vectors of $\Upsilon$ and $\gamma$, respectively. The unique helicity projector reads
\begin{equation}\label{helicity-etac}
\mathcal{P}_{0,1}^{(\eta_c)\mu \nu}=\frac{i}{2 m_{\Upsilon}|\mathbf{P}|} \epsilon^{\mu \nu \rho \sigma} Q_{\rho} P_{\sigma},
\end{equation}
which can be derived from the angular momentum conservation and parity conservation.
Thus the helicity amplitude can be extracted by acting the helicity projector on the amputated amplitude $\mathcal{A}^{(\eta_c)}_{0,1}=\mathcal{P}_{0,1}^{(\eta_c)\mu \nu}\mathcal{A}_{\mu \nu}^{(\eta_c)}$.

Very similarly, the amplitude $\mathcal{A}^{(\chi_{c0})}$ of the process $\Upsilon \to \chi_{c0}+\gamma$ can be expressed as
$\mathcal{A}^{(\chi_{c0})}=\mathcal{A}_{\mu \nu}^{(\chi_{c0})} \epsilon_{\Upsilon}^{\mu} \epsilon_{\gamma}^{* \nu}(\lambda_{2})$.
There is only one helicity projector corresponding to $\lambda_1=0$ and $\lambda_2=1$, which can be expressed as
\begin{equation}
\mathcal{P}_{0,1}^{(\chi_{c0})\mu \nu}=-\frac{1}{2}g_{\perp}^{\mu\nu},
\end{equation}
which is quite different from the form of (\ref{helicity-etac}) because the parity of $\chi_{c0}$ is adverse to that of $\eta_c$.
The helicity amplitude of $\chi_{c0}$ can be deduced by acting the helicity projector on the amputated amplitude $\mathcal{A}^{(\chi_{c0})}_{0,1}=\mathcal{P}_{0,1}^{(\chi_{c0})\mu \nu}\mathcal{A}_{\mu \nu}^{(\chi_{c0})}$.

By parameterizing the amplitude $\mathcal{A}^{(\chi_{c1})}$ of the process $\Upsilon \to \chi_{c1}+\gamma$ into
$\mathcal{A}^{(\chi_{c1})}=\mathcal{A}_{\mu \nu \alpha}^{(\chi_{c1})} \epsilon_{\Upsilon}^{\mu} \epsilon_{\gamma}^{* \nu}(\lambda_{2})\epsilon_{\chi_{c1}}^{*\alpha}(\lambda_1)$, we can deduce the two independent helicity projectors for $\chi_{c1}$
\begin{subequations}
\begin{eqnarray}
\mathcal{P}_{1,1}^{(\chi_{c1})\mu \nu \alpha}&=&\frac{-1}{2m_\Upsilon |\mathbf{P}|} L_\Upsilon^{\mu}\epsilon^{\nu\alpha\rho \sigma}
Q_{\rho} P_{\sigma},\\
\mathcal{P}_{0,1}^{(\chi_{c1})\mu \nu \alpha}&=&\frac{1}{2m_\Upsilon |\mathbf{P}|} L_{\chi_{c1}}^{\alpha}\epsilon^{\mu\nu\rho \sigma}
Q_{\rho} P_{\sigma}.
\end{eqnarray}
\end{subequations}
The corresponding two helicity amplitudes of $\chi_{c1}$ can be readily obtained by acting the helicity projectors on the amputated amplitude: $\mathcal{A}^{(\chi_{c1})}_{0,1}=\mathcal{P}_{0,1}^{(\chi_{c1})\mu \nu\alpha}\mathcal{A}_{\mu \nu\alpha}^{(\chi_{c1})}$
and $\mathcal{A}^{(\chi_{c1})}_{1,1}=\mathcal{P}_{1,1}^{(\chi_{c1})\mu \nu\alpha}\mathcal{A}_{\mu \nu\alpha}^{(\chi_{c1})}$.

Finally, if we express the amplitude $\mathcal{A}^{(\chi_{c2})}$ of the process $\Upsilon \to \chi_{c2}+\gamma$ into
$\mathcal{A}^{(\chi_{c2})}=\mathcal{A}_{\mu \nu \alpha\beta}^{(\chi_{c2})} \epsilon_{\Upsilon}^{\mu} \epsilon_{\gamma}^{* \nu}(\lambda_{2})\epsilon_{\chi_{c2}}^{*\alpha\beta}(\lambda_1)$, the three independent helicity projectors for $\chi_{c2}$
can be deduced
\begin{subequations}
\begin{eqnarray}
\mathcal{P}_{2,1}^{(\chi_{c2})\mu \nu\alpha\beta}&=&
\frac{1}{4}\bigg(g_{\perp}^{\mu\nu}g_{\perp}^{\alpha\beta}-g_{\perp}^{\mu\alpha}g_{\perp}^{\nu\beta}
-g_{\perp}^{\mu\beta}g_{\perp}^{\nu\alpha}\bigg),\\
\mathcal{P}_{1,1}^{(\chi_{c2})\mu \nu\alpha\beta}&=&
\frac{-1}{2\sqrt{2}}L_\Upsilon^\mu\bigg(g_{\perp}^{\nu\alpha}L_{\chi_{c2}}^\beta+g_{\perp}^{\nu\beta}L_{\chi_{c2}}^\alpha\bigg),\\
\mathcal{P}_{0,1}^{(\chi_{c2})\mu \nu\alpha\beta}&=&
\frac{-1}{2\sqrt{6}}g_{\perp}^{\mu\nu}\bigg(g_{\perp}^{\alpha\beta}+2L_{\chi_{c2}}^\alpha L_{\chi_{c2}}^\beta\bigg).
\end{eqnarray}
\end{subequations}
The corresponding helicity amplitudes for $\chi_{c2}$ can be readily obtained by acting the helicity projectors on the amputated amplitude: $\mathcal{A}^{(\chi_{c2})}_{2,1}=\mathcal{P}_{2,1}^{(\chi_{c2})\mu \nu\alpha\beta}\mathcal{A}_{\mu \nu\alpha\beta}^{(\chi_{c2})}$, $\mathcal{A}^{(\chi_{c2})}_{1,1}=\mathcal{P}_{1,1}^{(\chi_{c2})\mu \nu\alpha\beta}\mathcal{A}_{\mu \nu\alpha\beta}^{(\chi_{c2})}$
and $\mathcal{A}^{(\chi_{c2})}_{0,1}=\mathcal{P}_{0,1}^{(\chi_{c2})\mu \nu\alpha\beta}\mathcal{A}_{\mu \nu\alpha\beta}^{(\chi_{c2})}$.

Since we do not consider the relativistic corrections in this work, it is eligible to make the approximations $m_\Upsilon\approx 2m_b$, $m_{\eta_c}\approx 2m_c$, and $m_{\chi_{cJ}}\approx 2m_c$ in the helicity projectors.

\section{Applying the Cheng-Wu theorem to deal with loop integrals~\label{appendix-Cheng-Wu}}

The description about Cheng-Wu theorem can be found in Refs.~\cite{Cheng-Wu-1,Bjoerkevoll:1992cu,Smirnov_Book,Jantzen:2012mw}.
In Ref.~\cite{Sang:2020zdv}, the authors have successfully gotten rid of the nontrivial Heaviside step function through Cheng-Wu theorem.
In Ref.~\cite{Feng:2021kha}, one of our authors has applied the Cheng-Wu theorem to evaluate
some special integrals. In this appendix, we further extend the application of Cheng-Wu theorem in multi-loop
integration.

The function ${\mathcal F}$ is projective, if it satisfies
\begin{equation}
{\cal F}(\lambda x_1,\cdots,\lambda x_n) = \lambda^{-n} {\mathcal F}(\lambda x_1,\cdots,\lambda x_n).
\end{equation}
The Cheng-Wu theorem indicates that the integral ${\mathcal I}$
\begin{eqnarray}\label{eq-Cheng-Wu-I-1}
	{\mathcal I} &=& \int_0^\infty d x_1 \cdots \int_0^\infty d x_n \, {\cal F}(x_1,\cdots,x_n) \delta(1-\sum_{i=1}^n a_i x_i)
\end{eqnarray}
does not depend on the set \{$a_i$\}, when all variables in \{$a_i$\} are nonnegative and at least one $a_i$
is positive, i.e., $\forall i, a_i\ge0$ and $\exists k, a_k>0$.
The theorem might be employed to eliminate the
negative terms in ${\mathcal F}$.

After the alpha-representation, the Feynman multi-loop integrals can be generally transferred into the following
form
\begin{equation}\label{eq-Cheng-Wu-I-2}
{\mathcal I} = \int_0^\infty d{\bm x} \; \delta(1-\sum_{i=1}^n x_i)\; U({\bm x})^a F({\bm x})^b,
\end{equation}
where ${\bm x}$ denotes the set \{$x_i$\}, and the exponents $a$ and $b$ are linear in $\epsilon$ with the dimension of the spacetime $d=4-2\epsilon$.
It can be proved that eq.~(\ref{eq-Cheng-Wu-I-2}) conforms to the form of (\ref{eq-Cheng-Wu-I-1}).
The $U$ term and $F$ term are the polynomials of ${\bm x}$. In the physical region, the $F$ term may contain negative terms.
In the following, we describe the strategy to eliminate the negative terms in $F$ by employing the Cheng-Wu theorem (\ref{eq-Cheng-Wu-I-1}).
Since the $U$ term is irrelevant to the discussion, we will suppress the $U$ term in the remainder of this appendix.

We introduce how to eliminate the negative terms in $F$ through three specific cases.

{\it Case} 1: If $F$ can be expressed into $F={\cal P}(\hat{\bm{x}}_k)-x_k {\cal Q}(\hat{\bm{x}}_k)-i\varepsilon$, where
$\hat{\bm{x}}_k$ denotes the set \{$x_i$\} with the variable $x_k$ removed,
and ${\cal P}$ and ${\cal Q}$ are positive polynomials of $\hat{\bm{x}}_k$.
We are able to pull $x_k$ outside the $\delta$-function through setting $a_k=0$ in eq.~(\ref{eq-Cheng-Wu-I-1}), and then
rescale $x_k$ through
\begin{equation}
x_k = \frac{{\cal P}(\hat{\bm{x}}_k)}{{\cal Q}(\hat{\bm{x}}_k)} \frac{y_k}{x_j},
\end{equation}
where we call $x_j$ the label variable which should be different from $x_k$.
After renaming $y_k$ with $x_k$ and putting $x_k$ back inside $\delta$-function by the Cheng-Wu theorem,
we finally obatin
\begin{equation}
{\cal I} = \int_0^\infty d{\bm{x}} \; \delta(1-\sum_{i=1}^n x_i) \; \frac{{\cal P}^{b+1}}{x_j^{b+1}\, {\cal Q}} \; \; \left( x_j-x_k-i\varepsilon\right)^{b},
\end{equation}
where we have suppressed the arguments of ${\cal P}$ and ${\cal Q}$ for simplification. After the transformation, the new $F$
term turns to be $x_j-x_k-i\varepsilon$.
Splitting the integration domain into two parts $x_j\le x_k$ and $x_j\ge x_k$ through the transformations
\begin{equation}
x_k = x_j + x_k
\end{equation}
and
\begin{equation}
x_j = x_j + x_k,
\end{equation}
we can convert the $F$ term either into a fully positive polynomial or into a fully negative polynomial, where the $F$ term can be
further turned to be positive by an overall constant phase factor for the latter case.

{\it Case} 2: If $F$ can be expressed into $F={\cal P}(\hat{\bm{x}}_{kl})-x_k {\cal Q}_k(\hat{\bm{x}}_{kl})-x_l {\cal Q}_l(\hat{\bm{x}}_l)-i\varepsilon$, where $\hat{\bm{x}}_{kl}$ denotes the set \{$x_i$\} with both $x_k$ and $x_l$ removed, and
${\cal P}$, ${\cal Q}_{k}$ and ${\cal Q}_{l}$ are positive polynomials.
We can first pull $x_k$ and $x_l$ outside the $\delta$-function, and then rescale the two variables
\begin{equation}
x_l = \frac{{\cal P}(\hat{\bm{x}}_{kl})}{{\cal Q}_l(\hat{\bm{x}}_l)} \frac{y_l}{x_j},
\quad
x_k = \frac{{\cal P}(\hat{\bm{x}}_{kl})}{{\cal Q}_k(\hat{\bm{x}}_{kl})} \frac{y_k}{x_j},
\end{equation}
where we call $x_j$ the label variable which should be different from $x_k$ and $x_l$.
After renaming $y_k$ and $y_l$ with $x_k$ and $x_l$ respectively, and putting $x_k$ and $x_l$
back inside the $\delta$-function by the Cheng-Wu theorem,
we then obtain
\begin{equation}
{\cal I} = \int_0^\infty d\bm{x} \, \delta(1-\sum_i x_i) \, \frac{{\cal P}(\hat{\bm{x}}_k)^{b+2}}{x_j^{b+2}{\cal Q}_l(\hat{\bm{x}}_l){\cal Q}_{k}(\hat{\bm{x}}_{kl})} \, \left( x_j-x_k-x_l -i\varepsilon \right)^{b}.
\end{equation}
Now the new $F$ term becomes $x_j-x_k-x_l -i\varepsilon$.
Once again, by pulling $x_j$ outside the $\delta$-function, rescaling $x_j$ according to
\begin{equation}
x_j = (x_k+x_l)\frac{y_j}{x_k},
\end{equation}
renaming $y_j$ with $x_j$, and putting $x_j$ back inside the $\delta$-function, we finally arrive at
\begin{equation}
{\cal I} = \int_0^\infty d\bm{x} \, \delta(1-\sum_{i} x_i) \, \frac{{\cal P}^{b+2}}{x_j^{b+2}} \, \frac{x_k}{x_k+x_l} \, \frac{1}{{\cal Q}_l} \, \frac{1}{{\cal Q}_{k}} \, \left( x_j-x_k -i\epsilon \right)^{b}.
\end{equation}
 We can further divide the integration domain into two parts $x_j\le x_k$ and $x_j\ge x_k$ through the transformations
\begin{equation}
 x_k = x_j + x_k,
\end{equation}
and
\begin{equation}
x_j = x_j + x_k.
\end{equation}
We end up with either a positive $F$ polynomial or a fully negative $F$ polynomial which can be
turned to be positive by an overall constant phase factor.

{\it Case} 3: If $F$ can be expressed into $F={\cal P}(\hat{\bm{x}}_{kl})-x_k [{\cal Q}_k(\hat{\bm{x}}_{kl})-x_l {\cal Q}_l(\hat{\bm{x}}_{kl})]-i\varepsilon$, where ${\cal P}$, ${\cal Q}_{k}$ and ${\cal Q}_{l}$ are positive polynomials.
We first pull $x_l$ outside the $\delta$-function, and then rescale $x_l$ by
\begin{equation}
x_l=\frac{{\cal Q}_k(\hat{\bm{x}}_{kl})}{{\cal Q}_l(\hat{\bm{x}}_{kl})}\frac{y_l}{x_j},
\end{equation}
where $x_j$ is called label variable which should be different from $x_k$ and $x_l$.
By repeating the same procedure, we are able to put $x_l$ back inside $\delta$-function.
After the transformation,
the $F$ term now becomes ${\cal P}(\hat{\bm{x}}_{kl})-x_k\frac{{\cal Q}_k(\hat{\bm{x}}_{kl})}{x_j}(x_j-x_l)-i\varepsilon$.
The integration domain can be divided into two parts: $x_j\le x_l$ through the transformation
\begin{equation}
x_l = x_j + x_l ,
\end{equation}
and $x_j\ge x_l$ through
\begin{equation}
x_j = x_j + x_l.
\end{equation}
In the domain $x_j \le x_l$, the $F$ term is obvious positive.
For $x_j \ge x_l$, the $F$ term is of the same form as that in {\it Case} 1, thus we can repeat the procedure
in {\it Case} 1 to transfer the $F$ term positive.

Through the Cheng-Wu theorem, we can convert most of the complex-valued MIs encountered in this work into the real-valued MIs, which
makes the numerical integration much efficient.

\end{document}